\documentclass[aps,prb,twocolumn,amsmath,amssymb,superscriptaddress]{revtex4}
\usepackage{epsfig}
\usepackage{graphicx}
\usepackage{dcolumn}
\usepackage{hyperref}
\usepackage{ulem}
\usepackage{bm}
\usepackage{color}
\usepackage{braket}
\usepackage{latexsym}
\usepackage{amsmath}
\usepackage{amssymb}
\hypersetup{bookmarks,colorlinks,linkcolor = black,citecolor=blue,filecolor=blue,urlcolor=blue}
\usepackage{latexsym}

\def \beq{\begin{eqnarray}}
\def \eeq{\end{eqnarray}}

\def \B{{\bm B}}

\def \bS{{\bm S}}
\def \r{{\bm r}}
\def \rp{{\bm r^{\prime}}}
\def \bro{{\bm\rho}}
\def \m{{\bm m}}
\def \n{{\bm n}}
\def \k{{\bm k}}
\def \q{{\bm q}}
\def \N{{\mathcal N}}
\def \S{{\mathcal S}}
\def \T{{\mathcal{T}}}
\def \on{{\omega_n}}

\def \tz{{\text{z}}}
\newcommand{\nn}{\nonumber \\}
\newcommand{\sgn}{\text{sign}}

\newcommand{\nocontentsline}[3]{}
\newcommand{\tocless}[2]{\bgroup\let\addcontentsline=\nocontentsline#1{#2}\egroup}

\begin{document}
\title{{\textbf{Diagnosing phases of magnetic insulators via noise magnetometry with spin qubits}}}
\author{Shubhayu Chatterjee}
\affiliation{Department of Physics, Harvard University, Cambridge Massachusetts
02138, USA.}
\affiliation{Department of Physics, University of California, Berkeley, California 94720, USA.}
\author{Joaquin F. Rodriguez-Nieva}
\affiliation{Department of Physics, Harvard University, Cambridge Massachusetts
02138, USA.}
\author{Eugene Demler}
\affiliation{Department of Physics, Harvard University, Cambridge Massachusetts
02138, USA.}


\begin{abstract}

Two-dimensional magnetic insulators exhibit a plethora of competing ground states, such as ordered (anti)ferromagnets,  exotic quantum spin liquid states with topological order and anyonic excitations, and random singlet phases  emerging in highly disordered frustrated magnets. Here we show how single spin qubits, which interact directly with the low-energy excitations of magnetic insulators, can be used as a diagnostic of magnetic ground states. Experimentally tunable parameters, such as qubit level splitting, sample temperature, and qubit-sample distance, can be used to measure spin correlations with energy and wavevector resolution. Such resolution can be exploited, for instance, to distinguish between fractionalized excitations in spin liquids and spin waves in magnetically ordered states, or to detect anyonic statistics in gapped systems. 

\end{abstract}

\maketitle


\section{Introduction}

The subtle interplay between strong correlations, geometric or exchange frustration, disorder and quantum fluctuations in insulators with spin degrees of freedom can lead to a variety of ground states that often compete closely in energy.\cite{SS_QMagReview,Balents2010,SS_QMagNatPhys} The most common phases exhibit long range magnetic order which spontaneously break the underlying spin-rotation symmetry of the Hamiltonian. Alternatively, strong quantum fluctuations in lower dimensional systems can lead to exotic quantum spin liquid (QSL) phases, which are characterized by intrinsic topological order and anyonic excitations described by lattice gauge theories.\cite{Lee,SavaryBalentsQSLReview} Another possible ground state is the valence bond solid (VBS), which preserves spin-rotation symmetries but breaks the discrete translation symmetry of the crystal.\cite{SB1990,Uhrig2004} In the presence of strong disorder in the exchange coupling between neighboring spins, the VBS can form a random singlet phase which statistically preserves all symmetries, but is topologically trivial.\cite{Itamar1,Itamar2} Given the wide spectrum of possibilities, 
it is of primary importance to develop experimental probes that can distinguish between these competing ground states, and find convincing signatures of their corresponding emergent collective excitations. 

The recent introduction of single spin qubits, such as
Nitrogen Vacancy (NV) centers  
in diamond,\cite{nvreview} as nanoscale probes of correlated materials enables new pathways to access the physics of magnetic insulators. Optical initialization and read-out capabilities of their spin states, precise manipulations by resonant microwave pulses, efficient coupling to local magnetic fields, and excellent spatial resolution,  
make spin probes 
an ideal tool to probe both statics and dynamics of magnetic systems. Since the Zeeman splitting of the spin qubit can be measured optically with great accuracy, spin probes can be used to image local magnetic textures, even those induced by a single spin.\cite{nvsinglespin} Furthermore, the spin relaxation time induced by intrinsic fluctuations in a material can be used to probe charge and spin dynamics. For instance, the relaxation time can be used as a diagnostic of different regimes of electronic transport, ranging from ballistic to diffusive to hydrodynamic,\cite{Agarwal2017} spin-charge separation in one-dimensional systems,\cite{Joaquin18} and magnetic monopoles in spin-ice materials.\cite{Kirschner2018} In metallic states, noise is dominated by transverse fluctuations of charge currents, provided the system is not extremely localized.
\cite{Agarwal2017} However, in an insulator with a large gap to charged excitations, the magnetic noise is dominated by spin fluctuations. Thus, spin qubits can serve as a novel probe to distinguish between different competing ground states in insulating materials. 

\begin{figure}
  \centering\includegraphics[scale=1.0]{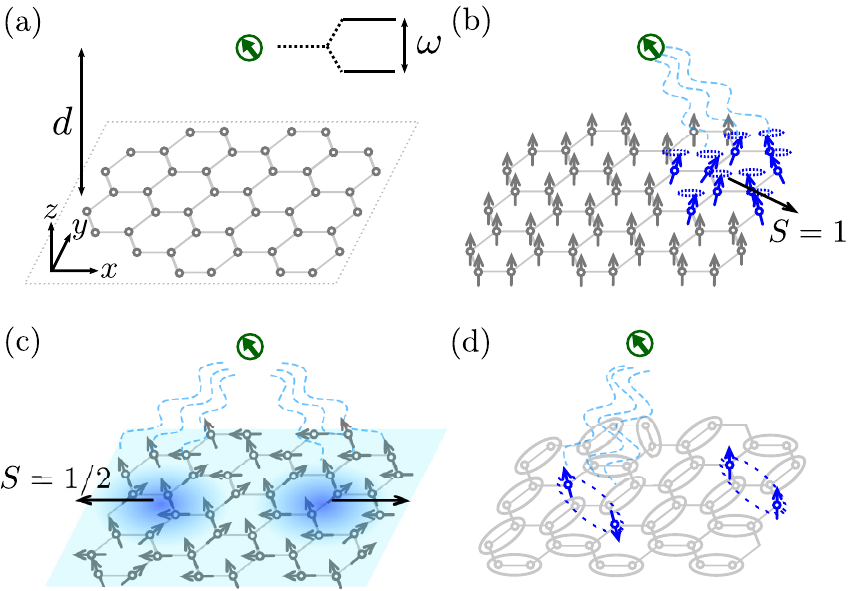}
  \caption{(a) Experimental setup showing a spin qubit located at a distance $d$ from a two-dimensional magnetic insulator. (b) Detection of low energy $S=1$ excitations in magnetically-ordered phases using a single spin qubit. (c) Schematic depiction of the detection of $S = 1/2$ fractionalized excitations in spin liquids. (d) Detection of low energy excitations in disordered valence bond solids.}
  \label{fig:schematics}
\end{figure}

In the present work,
we find the characteristic signatures of the underlying magnetic ground state on the spin qubit relaxation time. By tuning experimental parameters, 
we show how such signatures can be exploited to diagnose ground states.  
The time-scale for the relaxation 
of a spin qubit 
with level splitting 
$\omega$ depends on the magnetic noise spectrum of an insulator, which in turn is related to the spin-spin (retarded) correlation function 
\beq
{\cal C}_{\alpha\beta}(i,j,\omega) = -i\int_{0}^{\infty} dt e^{i\omega t}\langle [{S}_i^\alpha(t), {S}_j^\beta(0)]\rangle,
\label{eq:structurefactor}
\eeq
where $\langle ... \rangle$ is a short-hand notation 
for ensemble average.
In magnetically ordered states, ${\cal C}_{\alpha\beta}$ is dominated by gapless single-particle collective modes called spin-waves, or magnons, which are the $S=1$ Goldstone bosons of the spontaneously broken spin-rotation symmetry, see Fig.\ref{fig:schematics}(a)-(b). In quantum spin liquid (QSL) phases, the excitations carry fractional quantum numbers corresponding to the global symmetries of the Hamiltonian. For example, the spin-carrying excitations are $S=1/2$ spinons, each of which may be understood as `half a magnon'. While these can be created only in pairs by local operators, they can propagate as independent collective modes and therefore lead to a broad two-particle continuum in the dynamic spin structure factor, see Fig.\ref{fig:schematics}(c). This is distinct from 
the sharp peak that is seen for single particle excitations such as 
magnons. Finally, in a clean valence bond solid (VBS) state, the excitations are gapped $S=1$ triplons --- gapless Goldstone modes are absent as the relevant broken symmetry (i.e., lattice translation) is discrete. In a random singlet phase, which is the theoretically proposed fate of VBS phases in highly frustrated inorganic insulators in presence of disorder,\cite{Itamar1} the elementary excitations are gapped. However, the system appears gapless as the there is a distribution of low-energy levels, induced by pairing of unbonded spins [Fig.\ref{fig:schematics}(d)] which scales as a power law of energy for sufficiently large samples. 

The emergent excitations for gapped spin liquids may have anyonic statistics which have been difficult to detect in traditional settings. Inspired by the recent proposal\cite{Sid_PRL2017} to use threshold spectroscopy to detect anyonic statisics, 
we also outline how magnetic noise spectroscopy via spin probes, with its excellent energy and spatial resolution, 
can provide convincing signatures of non-trivial braiding statistics.

Importantly, spin qubits offer several significant advantages over conventional experimental probes of solid state systems. As we show explicitly below, the spin qubit is sensitive to the magnetic noise at wave-vectors $q \sim d^{-1}$ [$d$ being the sample-probe distance, see Fig.\ref{fig:schematics}(a)] and frequency  $\omega$ which is the level splitting of the qubit. Thus, by using both distance and transition frequency as tuning parameters, the dynamic spin structure factor can be measured with energy (up to several mK) and momentum resolution (up to a few nm). One major issue with most probes is that the physical observable they measure depend upon responses from multiple parts of the system, which can be difficult to isolate from one another. For example, the neutron-scattering cross-section and specific heat measurements in insulators depend on the cumulative contributions from spin-excitations and phonons. Single spin qubits bypass the problem by detecting spin-fluctuations directly without contamination from phonons. Spin qubits do not require the sample to be placed in a magnetic field for measurements, and therefore are not resolution-limited by magnetic field gradients unlike NMR. Further, because they are point-like probes with nanometer resolution, they have the potential to bridge the large length-scale gap between scanning tunneling microscopy and global transport or thermodynamic susceptibility measurements. In addition, as the spin probe does not require a driving field, it is minimally invasive. This is not generally true for transport probes that distinguish different magnetic states;\cite{SCSS15,WCMB17,BBK_Science} these run the risk of driving the system into non-linear responses via external perturbing fields, making the results challenging to interpret.
Quite a few probes which have been suggested to provide smoking gun evidence for exotic states in quantum magnetism have significant experimental hurdles to their realization.\cite{SavaryBalentsQSLReview} On the contrary, spin qubits 
are currently being used to measure local magnetic textures\cite{Tetienne2015,Van2015,Dovzhenko2018}, spin chemical potentials\cite{Du2017} and ferromagnetic phase transitions in metals \cite{HsiehNV2018} over a wide range of physical parameters like temperature/pressure. As a result, they hold great promise for detection of novel phases in insulating magnets, particularly in layered quasi two-dimensional materials or the surfaces of three dimensional materials. 

The rest of the paper is organized as follows. In Sec.~\ref{sec:RT}, we develop the general formalism for noise magnetometry in insulating two-dimensional states, and explicitly compute 
the 
relaxation time-scale $T_1$ as a function of  spin-correlation functions. 
In Sec.~\ref{sec:gapless}, we apply the formalism to magnetically ordered states, quantum spin liquids and clean/disordered VBS states, and discuss their salient features which can be used to pinpoint the ground state in a given material. In Sec.~\ref{sec:anyons}, we derive the dependence of the relaxation time on the anyonic statistics in gapped systems. In Sec.~\ref{sec:materials}, we discuss the implications of our results for promising material candidates for the different phases. In Sec.~\ref{sec:conc}, we summarize our main results. The Appendices contain the details of the calculations. 
\section{General formalism for relaxation time}
\label{sec:RT}

We start by developing a general formalism to relate the relaxation time $T_1$ of a spin qubit induced by the magnetic noise generated by a two-dimensional insulating sample. This treatment closely follows the corresponding formalisms for two-dimensional metals and one-dimensional Luttinger liquids.\cite{Agarwal2017,Joaquin18} 

We consider the spin qubit placed at $\r_{\rm q} = (0,0,d)$, above the two dimensional insulator on the $x$-$y$ plane, as shown in Fig.~\ref{fig:schematics}. The spin probe can be treated as a two-level system with an intrinsic level-splitting of $\omega_0$, which can be varied by a static Zeeman field $\bm{B}_0$.  
The Hamiltonian of the combined probe and magnet system is given by
\beq
\mathcal{H} = {\cal H}_{\rm q} + {\cal H}_{\rm q-m} + {\cal H}_{\rm m}.
\eeq
The term ${\cal H}_{\rm q}$ is the spin qubit Hamiltonian ($\hbar = 1$),
\beq
{\cal H}_{\rm q} = \frac{\omega}{2} \hat{\bm n}_{\rm q}\cdot{\boldsymbol\sigma},
\eeq 
where $\hat{\bm n}_{\rm q}$ is the unit vector along the direction of $\omega_0 \hat{\bm n}^\prime_{\rm q} + \bm{B}_0$, and $\hat{\bm n}^\prime_{\rm q}$ is the direction of the intrinsic polarizing field of the qubit. 
For instance, in the case of NV centers in diamond, $\hat{\bm n}^\prime_{\rm q}$ is the axis of the NV defect in the diamond lattice. Thus, $\omega = (\omega_0^2 + B_0^2 + 2 \omega_0 \mathbf{B}_0 \cdot \hat{\mathbf{ n}}^\prime_{\rm q})^{1/2}$ is the resulting probing frequency. 
The term ${\cal H}_{\rm m}$ is the Hamiltonian of the two-dimensional magnetic sample which will be specified below for different ground states. Finally, the term ${\cal H}_{\rm q-m}$ is the qubit-magnet coupling induced by dipole-dipole interactions: 
\beq
{\cal H}_{\rm q-m} = \mu_{\rm B}{\boldsymbol\sigma}\cdot{\bm B}, \,\,\, {\bm B} =\frac{\mu_0\mu_{\rm B}}{4\pi} \sum_j \left[\frac{{\bm S}_j}{r_j^3} - \frac{3({\bm S}_j\cdot{\bm r}_j) {\bm r}_j}{r_j^5}\right].
\label{eq:Bfield}
\eeq
Here ${\bm B}$ is the time-dependent magnetic field at the position of the probe induced by spin fluctuations 
in the 2D magnet, and ${\bm r}_j = (x_j,y_j,-d)$ is the relative position between the $j$-th spin in the two-dimensional lattice and 
the probe. 

The relaxation time of the qubit can be related to the retarded correlators of the fluctuating magnetic field arising from the sample via Fermi's Golden rule and the fluctuation-dissipation theorem. 
In thermal equilibrium at temperature $T$, the 2D insulator 
is described by the density matrix $\rho = \sum_n \rho_n|n\rangle\langle n|$, with $|n\rangle$ the eigenstates of ${\cal H}_{\rm m}$ with energy $\varepsilon_m$, 
and $\rho_n = e^{-\varepsilon_n/T}$. 
The absorption rate, $1/T_{\rm abs}$, and emission rate, $1/T_{\rm em}$, is obtained from Fermi Golden's rule using the initial state $|i\rangle = |-\rangle_{\rm q}\otimes\rho$ and $|i\rangle = |+\rangle_{\rm q}\otimes\rho$, respectively (for $\hat{\bm n}_{\rm q} = \hat{z}$): 
\beq
1/T_{\rm abs,em} = 2\pi \sum_{nm} \rho_n {B}_{nm}^\pm{B}_{mn}^\mp \delta (\omega \pm \varepsilon_{mn}),
\eeq
where ${B}_{nm}^\alpha = \langle n | {\hat B}^\alpha | m \rangle$, $B^\pm = B^x \pm i B^y$, and $\varepsilon_{mn}$ is the energy difference between states $m$ and $n$, $\varepsilon_{mn} = \varepsilon_m - \varepsilon_n$. The relaxation rate, defined as 
$1/T_{1} = [1/T_{\rm abs} + 1/T_{\rm em}]/2$,  
can be expressed as
\beq
\frac{1}{T_1} =\frac{(\mu_0 \mu_{\rm B})^2}{2} \int_{-\infty}^{\infty} dt e^{i \omega t} \langle \{ B^-(t), B^+(0) \} \rangle, 
\eeq
where $\{,\}$ denotes anticommutation. 
Using the fluctuation-dissipation theorem, $1/T_1$ can be expressed in terms of the retarded correlation function as 
\beq
\begin{array}{c}
\displaystyle \frac{1}{T_1} =\frac{(\mu_0 \mu_{\rm B})^2}{2} {\rm coth}\left(\frac{\omega}{2T}\right) \left(- {\rm Im}\left[C^R_{B^- B^+}(\omega)\right] \right), \\ \\
\displaystyle C^R_{B^\alpha B^\beta}(\omega) = -i \int_{0}^{\infty} dt e^{i \omega t} \langle [ B^\alpha(\r,t), B^\beta(\r,0) ] \rangle. 
\end{array}
\label{eq:T1}
\eeq

Finally, $1/T_1$ can be expressed in terms of spin-spin correlation functions by  
inserting Eq.(\ref{eq:Bfield}) into Eq.(\ref{eq:T1}) and going into momentum space ($a = $ lattice spacing): 
\beq
\begin{array}{rl}
\displaystyle \frac{1}{T_1} = & \displaystyle \frac{\mu_0^2 \mu_{\rm B}^4}{8a^2} {\rm coth}\left(\frac{\omega}{2T}\right) \int \frac{d^2\q}{(2\pi)^2} e^{-2 q d} q^2 \left[{\cal C}_{-+}''({\bm q},\omega) \right. 
\\ \\  
& \displaystyle \left. +{\cal C}_{+-}''({\bm q},\omega)+4{\cal C}_{zz}''({\bm q},\omega)\right], 
\end{array}
\label{eq:T12}
\eeq
which is the central result of this section. Here, ${\cal C}_{\alpha\beta}({\bm q},\omega) = \frac{1}{N}\sum_j e^{i{\bm q}\cdot({\bm r}_j-{\bm r}_i)} {\cal C}_{\alpha\beta}(i,j,\omega)$ is the spatial Fourier transform of the spin-correlation function defined in Eq.(\ref{eq:structurefactor}), where we have used its translational invariance ($N = $ number of lattice sites), and ${\cal C}^{''}_{\alpha\beta} = - {\rm Im}[{\cal C}_{\alpha\beta}]$.

The relaxation time has several experimentally tunable knobs which can be varied to provide valuable information about spin-spin correlations in the sample. Equation (\ref{eq:T12}) shows that the $\q$ integral has an argument of $q^3 e^{- 2 q d}$ originating from the dipole-dipole interaction represented in momentum space and the Jacobian for 2D integration, resulting in a filtering function which is peaked around $q \sim d^{-1}$. Such $d$ dependence allows to selectively probe ${\cal C}_{\alpha\beta}$ at different wavevectors. By the same token, it is also possible to vary $\omega$ with a static magnetic field at fixed $T$ and study the relaxation time at different energies  or, alternatively, study  the relaxation time as a function of temperature $T$ at fixed $\omega$. All of these furnish valuable information about the spin-correlations in the insulating sample with energy and momentum resolution. 

To illustrate how the relaxation time varies as a function of experimentally tunable parameters, 
we consider the simplest case of probing paramagnetic fluctuations 
at different values of
$d$. In this case, it is possible to relate $T_1$ 
 to the magnetic field created by spins within a lengthscale $D$ of the sample. Typically, $D \sim d$ is related to the sample-probe distance, but it can also be determined 
by other emergent lengthscales in the system (for instance, magnon momentum at frequency $\omega$, see below). The relaxation 
time is proportional to magnetic field fluctuations which are induced by magnetic dipoles, $B \sim \mu_0\mu_{\rm B} S_i / D^3$, which results in 
\beq
\begin{array}{rl}
\displaystyle \frac{1}{T_1} &  
\approx \mu_0^2\mu_{\rm B}^4\sum_{i,j} \langle[S_{i\alpha}/D^3,S_{j\alpha}/D^3]\rangle_\omega \\
& \displaystyle = \frac{\mu_0^2\mu_{\rm B}^4}{D^{6}} \int d^2\mathbf{R} \int d^2\r \langle[S_\alpha(\r),S_\alpha(0)]\rangle_\omega.
\end{array}
\eeq
In the last step, 
we have made a continuum approximation, and used translational  invariance of the spin correlations to separate the integration into center of mass and relative coordinates ($\mathbf{R}$ and $\r$ respectively), both of which are integrated over regions of linear dimensions $D$. For a trivial paramagnet (as well as gapped spin liquids), the spin-correlations 
decay exponentially with a correlation length  
of a few lattice spacings. Therefore, the integral over $\r$ 
gives a constant, the ${\bm R}$ integral gives a factor of $D^2$, 
resulting in a relaxation time that scales as $D^{-4}$ for paramagnetic insulators with a spin gap. 

Different power laws of $D$ are obtained in magnetic materials with power law correlations of the form $\langle[S_\alpha(\r),S_\alpha(0)]\rangle_\omega \propto 1/r^\delta$, such as gapless spin liquids. The precise exponent $\delta$ depends on the dispersion of the gapless excitations, presence of disorder and nature of gauge flutuations in the spin liquid phase. In this case, the integral over ${\bm r}$ results in $\int d^2\r \langle[S_\alpha(\r),S_\alpha(0)]\rangle_\omega \sim D^{2-\delta}$ and, hence, the relaxation time scales as $D^{-(2+\delta)}$ with distance. 

In other regimes, the lengthscale $D$ is emergent from the sample physics, and is related to the frequency dependence of the dynamic spin correlations. In magnetically ordered phases, the transverse spin correlations at low energies have a delta function of the type $\delta(\omega - v_s q^\gamma)$, where $\gamma = 2$ ($1$) for (anti)ferromagnets (and $v_s$ is the inverse effective mass/spin-wave velocity). This fixes a lengthscale $D = q^{-1} = (\omega/v_s)^{1/\gamma}$ that is tied to the probe frequency. In spin-liquid phases at low temperatures, the imaginary part of spin-correlations have a step function form $\theta(\omega - v q)$, which again define the relevant distance scale $D = v/\omega$ for low probe frequencies. In such cases, the power law in $D$ translates to power laws in the probe frequency $\omega$. All these regimes will be discussed case by case below. 

Our results are valid for a single two-dimensional layer of insulating magnetic material. In the case of a quasi-two dimensional material, as is relevant for several frustrated magnets, the weakly coupled layers within a distance $D$ will give rise to independent contributions that add incoherently, and therefore the observed power law will be $D^{-(1 + \delta)}$. The case of fully three-dimensional materials require us to go beyond the independent layers approximation, and is left for future work.

\section{Applications to gapless systems}
\label{sec:gapless}

In this section, we discuss the characteristic $1/T_1$ behavior for two-dimensional gapless phases, including magnetically ordered states with Goldstone modes, topologically ordered spin liquids with Dirac cones or Fermi surfaces, and dirty VBS phases with gapless spin defects that form random singlets. The phases discussed in the present section, with the exception of U(1) spin liquids, can be obtained from a spin Hamiltonian on the honeycomb lattice with local interactions: 
\beq
{\cal H}_{\rm m} = \sum_{\langle ij \rangle^\mu} \sum_{\alpha} J_{ij}^{\mu\alpha} \sigma^\alpha_i \sigma^\alpha_j + \sum_{i} \bm{B}_i \cdot \bm{\sigma}_i.
\label{eq:pH}
\eeq
This Hamiltonian is an extension of the Kitaev model\cite{Kitaev06} with additional perturbations and/or disorder. In Eq.(\ref{eq:pH}), the link connecting nearest neighbor sites in the direction $\mu = x,y,z$ is labeled as $\langle ij \rangle^\mu$ (see Fig.~\ref{fig:kitaev}), 
and $\bm{B}_i$ is
a local magnetic field. {\it A priori}, the magnetic field and all the couplings are allowed to vary spatially so that we can describe both the clean and dirty limits, which give rise to qualitative distinct behaviors. 

We emphasize that Eq.(\ref{eq:pH})  
is not a description of any particular candidate material. Instead, it is constructed for the sole purpose of describing 
magnetic phases 
in different limits. Importantly, 
the character (dispersion, statistics, etc) of low-energy spin-carrying excitations in continuous symmetry-breaking or topological ordered phases, which are responsible magnetic fluctuations, are robust and independent of the details of the parent Hamiltonian. 
Hence, ${\cal H}_{\rm m}$ in Eq.(\ref{eq:pH}) provides a convenient starting point to study magnetic noise in several phases.

\subsection{Magnetically ordered states}

Let us consider first the Heisenberg Hamiltonian with $J_{ij}^{\mu,\alpha} = -J_H < 0$, such that
\beq
{\cal H}_{\rm m} = -J_H \sum_{\langle ij \rangle} {\boldsymbol\sigma}_i \cdot {\boldsymbol\sigma}_j.
\eeq
Assuming that the system is in the ferromagnetically ordered phase below the critical temperature $T_{\rm c}$, and without loss of generality, we set 
the $z$-axis as the axis of spin polarization (which may be canted from the two-dimensional plane of the systems).
Defining the operators $\sigma_j^\pm = \sigma_j^x\pm i \sigma_j^y$, 
the Hamiltonian can be expressed as 
\beq
{\cal H}_{\rm m} = - \frac{J_H}{2} \sum_{\langle i,j \rangle} \left(\sigma_i^+\sigma_j^- + \sigma_i^-\sigma_j^+ + 2 \sigma_i^z\sigma_j^z\right) .
\eeq
The low energy, effective theory of the Heisenberg ferromagnet can be described in terms of bosonic degrees of freedom using the Holstein-Primakoff transformation for spin-$1/2$ operators, $\sigma_i^+ = a_i^\dagger\sqrt{1-a_i^\dagger a_i}$, $\sigma_i^- = \sqrt{1-a_i^\dagger a_i}a_i^\dagger$, and $\sigma_i^z = -1/2 + a_i^\dagger a_i$, where $[a_i,a_j^\dagger] = \delta_{i,j}$. To quadratic order in $a_i$ and $a_i^\dagger$, this results in the spin wave Hamiltonian ${\cal H}_{\rm m}\approx{\cal H}_{\rm H}$
\beq
{\cal H}_{\rm H} = - J_{H} \sum_{\langle i,j \rangle} a_i^\dagger a_j + J_H \sum_i a_i^\dagger a_i. 
\eeq
After taking Fourier transform, $a_{\bm k} = \frac{1}{\sqrt{N}}\sum_i e^{i{\bm k}\cdot{\bm r}_i}a_i$, and $a_{\bm k}^\dagger = \frac{1}{\sqrt{N}}\sum_i e^{-i{\bm k}\cdot{\bm r}_i}a_i^\dagger$, the Heisenberg Hamiltonian describing low-energy spin waves is obtained:
\beq
{\cal H}_{\rm H} = \frac{J_H}{2} \sum_{\bm k} \left(\gamma_0 - \gamma_{\bm k}\right)a_{\bm k}^\dagger a_{\bm k}, \quad \quad \gamma_{\bm k} = \sum_{j}e^{i{\bm k}\cdot{\boldsymbol\delta}_j}.
\eeq
Here ${\boldsymbol\delta}_j$ denotes the nearest-neighbor vectors in the honeycomb lattice, see Fig.\ref{fig:kitaev}. Interactions between spin waves are governed\cite{Mattisbook} by the coupling constant $Ja^2({\bm k}\cdot{\bm p})$, and are negligibly small when temperature or energy (and thus the momenta $\k$ and $\bm p$ of colliding particles) is small. As such, the spin correlation functions are determined within single-particle physics, 
\beq
{\cal C}^{~''}_{\pm\mp}({\q},\omega) = \delta(\omega\mp\varepsilon_{\q}), \quad \varepsilon_{\bm q} = \frac{{\q}^2}{2m}, \quad m = 2/3Ja^2,  
\label{eq:Cpmmagnon}
\eeq
where quadratic dispersion is valid for $qa \ll 1$. 
The correlator ${\cal C}_{zz}$, instead, is a four-point correlation function which probes magnon transport. Usually it takes a diffusive form ${\cal C}_{zz} \sim 1/(\omega + i D q^2)$ and gives a contribution much smaller than ${\cal C}_{\pm\mp}$ when $\omega$ lies in the spin-wave continuum. 

As a result, the relaxation time of a spin qubit in close proximity to a 2D ferromagnet 
is 
governed by emission/absorption 
of long wavelength magnons with energy $\omega$. In particular, the frequency, temperature and distance dependence of the relaxation time can be obtained from Eq.(\ref{eq:T12}) combined with Eq.(\ref{eq:Cpmmagnon}), which gives rise to a relaxation time given by
\beq
\frac{1}{T_1} = \frac{\mu_0^2\mu_{\rm B}^4}{36 \pi a^6} \frac{\omega}{J_{\rm H}^2}{\rm coth}\left(\frac{\omega}{2T}\right)e^{-2q_{\omega}d},
\label{eq:T1magnon}
\eeq
with $q_\omega$ the magnon wavevector $q_\omega = \sqrt{4\omega / 3 J_{\rm H}}$. 
Note that, when the wavelength of the magnon is larger than the probe-to-sample distance, $1/T_1$ is independent of distance. Otherwise, $1/T_1$ 
decays exponentially with $d$, 
with a characteristic length given by the inverse magnon wavevector.  

The behavior for $1/T_1$ for an antiferromagnet is qualitatively similar to that in Eq.(\ref{eq:T1magnon}), but with minor differences. First, the dispersion relation is linear with $\omega$, $q_\omega \sim \omega / J_{\rm H}a$. Second, the spin-spin correlator for the 
antiferromagnet 
acquires an extra factor of $q$ because spins are anti-aligned in the bipartite lattice. This extra factors lead to $1/T_1 \propto \frac{\omega^4}{J_{\rm H}^2}{\rm coth} \left(\frac{\omega}{2T}\right)e^{-2q_{\omega}d}$. 

\subsection{Gapless quantum spin liquids}
\label{subsec:QSL}

Quantum spin liquids are long range entangled states that lack long-range magnetic order, and possess excitations that carry fractional values of global symmetries such as spin-rotation.\cite{Lee,Balents2010,SavaryBalentsQSLReview} Since gapless fractionalized bosonic excitations would condense at low temperatures, here we are interested in spin models with emergent charge-neutral fermionic excitations. However, the local Hilbert space is bosonic, and this implies that individual fermionic excitations must be non-local and occur in pairs. Theoretical descriptions of spin liquids require the non-local fermionic excitations to be coupled to emergent gauge fields, which can be gapped $\mathbb{Z}_2$ [gapless U(1)] and mediate short-range (long-range) interactions between the fermions.\cite{WenSqLattice} For the sake of concreteness and better analytical control, we primarily focus on the $\mathbb{Z}_2$ spin liquids with low-energy spin-half fermionic excitations within the framework of the Kitaev honeycomb model\cite{Kitaev06} with added perturbations. However, our results depend only on the nature of low-energy excitations and, as a result, they are more general. We also comment on relaxation times for gapless U(1) spin liquids, which are theoretically less controlled due to strongly coupled gapless excitations in both gauge and matter sectors. In all cases, the relaxation times show qualitatively distinct behavior as a function of the probe frequency,  sample-probe distance and temperature. Our main results are summarized in the Tables~\ref{tab:gapless1} and \ref{tab:gapless2}. The details of several computations may be found in Appendix \ref{app:integrals2}. The physically applicable regime according to the current experimental capabilities is $\omega \ll T$ as typical spin probes operate at GHz frequencies which are roughly hundred times smaller than the typical operating temperatures $T \approx 4$ -- $300$\,K.\cite{Du2017} However, we consider both the $\omega \gg T$ and $\omega \ll T$ regimes, keeping in mind the possibility of lower temperatures or larger spin-probes level splitting $\omega$ in the future.

\tocless\subsubsection{Clean $\mathbb{Z}_2$ QSL with Dirac spinons}

\begin{figure}
  \centering\includegraphics[scale=0.4]{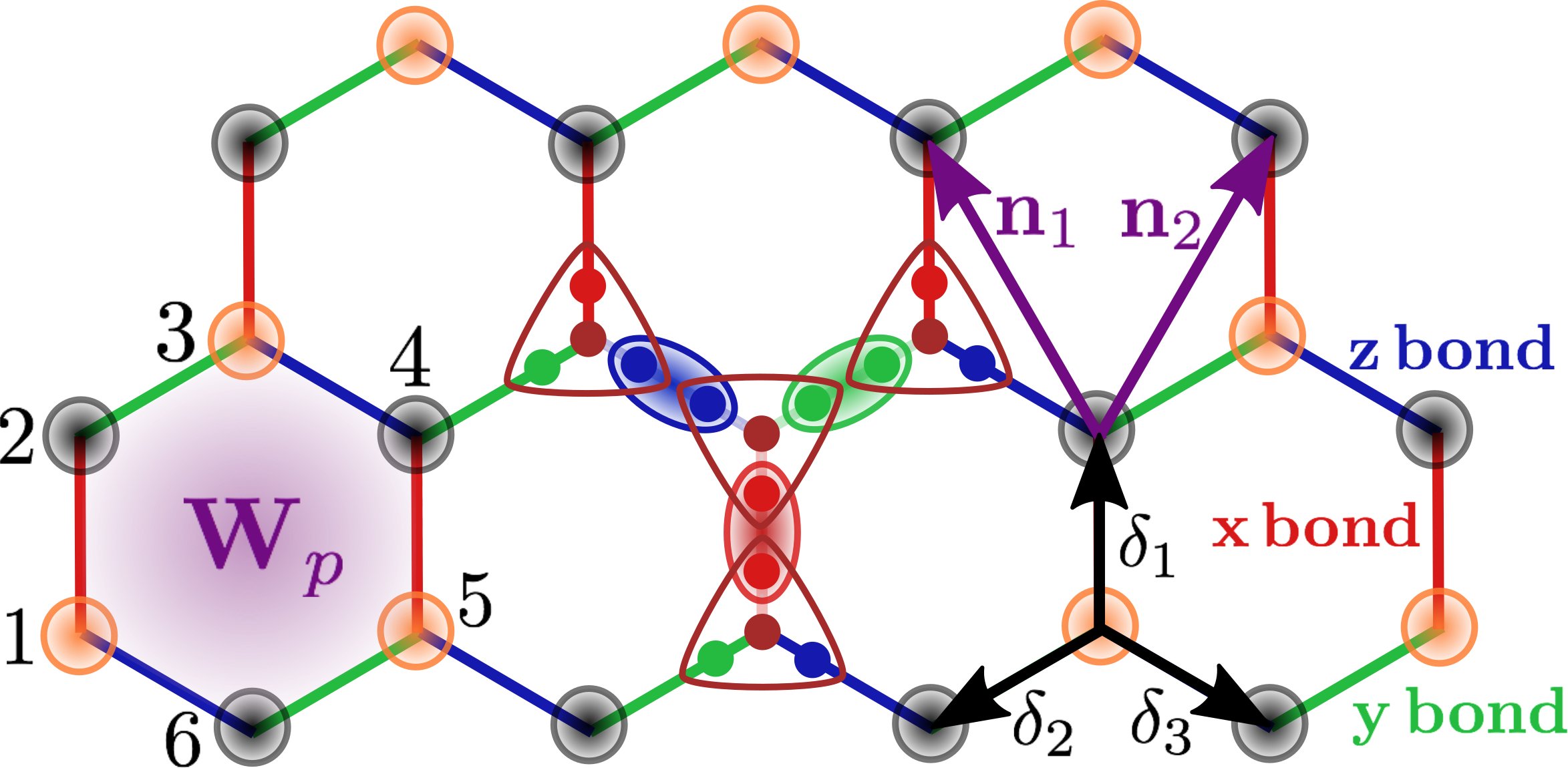}
  \caption{The honeycomb lattice of the Kitaev model. $\mathbf{n}_1$ and $\mathbf{n}_2$ are the two Bravais lattice vectors, and $\mathbf{\delta}_i$ $(i = 1,2,3)$ indicate nearest neighbors. The orange and gray circles correspond to the two sublattices. The rounded triangle at each site shows the 4 Majorana fermions used to write the Pauli spin operators. The free $c$ Majorana fermion is indicated by the brown dot. The red, green and blue dots refer to $b_i^x$, $b_i^y$ and $b_i^z$ Majorana fermions, and the corresponding bonds denote the bond variables $u_{ij}^{\mu}$ along the $x,y$ and $z$ links respectively.  $W_p$ is the flux operator defined in the main text, $W_p=-1$ correspond to gapped vison excitations.}
  \label{fig:kitaev}
\end{figure} 

We focus on the honeycomb spin liquid\cite{Kitaev06} beyond the Kitaev limit.\cite{SYB2016,KBM2018} The ground state of the pure Kitaev honeycomb model is a $\mathbb{Z}_2$ spin liquid and zero flux of the $\mathbb{Z}_2$ gauge field through the hexagonal plaquettes.\cite{Kitaev06} The emergent low-energy excitations are gapless fermions with a Dirac dispersion, and a gapped flux or vison. While the original honeycomb model has gapped spin correlations because a local spin operator necessarily creates a pair of gapped fluxes,\cite{GSR2007,SSKRG2011} it was shown in  Ref.~\onlinecite{SYB2016} that, in presence of symmetry-allowed perturbations expected to be present in material candidates,\cite{Trebst2017,KitaevSLReview} this gap disappears. As a result, the low-energy spectral weight from the emergent Dirac fermion has a major contribution to the dynamic spin structure factor at energies $\omega$ below the vison gap $\Delta_v$. 

We consider the Kitaev-Heisenberg-$\Gamma$ model,\cite{SYB2016,KBM2018,Jackeli,Chaloupka,KitaevSLReview} where the original Kitaev Hamiltonian $H_K$ is supplemented by Heisenberg and cross interactions: 
\beq
\begin{array}{rl}
{\cal H}_{\rm m} = & \displaystyle J_K \sum_{\langle ij \rangle^\mu} \sigma^\mu_i \sigma^\mu_j + \sum_{\mu (\nu \gamma)} \sum_{\langle ij \rangle^\mu} J_{\Gamma,\langle ij \rangle^\mu} (\sigma_i^\nu \sigma_j^\gamma + \sigma_i^\gamma \sigma_j^\nu) \\
& \displaystyle + J_H \sum_{\langle i,j\rangle}{\boldsymbol\sigma}_i\cdot{\boldsymbol\sigma}_j.
\end{array}
\label{eq:kitaevhgamma}
\eeq
Here $\nu, \gamma$ are the remaining two indices distinct from $\mu$, which is either $x,y$ or $z$ depending upon the direction of the link (see Fig.~\ref{fig:kitaev}). In the clean Dirac limit, we need to consider the Hamiltonian in Eq.~(\ref{eq:pH}), with translation invariant couplings $J_K$, $J_H$ and $J_\Gamma$ and no magnetic field $\bm\B = 0$.

We begin by recapitulating the Kitaev's exact solution to his original model. The bare Kitaev Hamiltonian $H_K$ is given by setting $J_\Gamma = J_{\rm H} = 0$ in Eq.~(\ref{eq:kitaevhgamma}), 
and where the bond connecting nearest neighbor sites in the direction $\mu = x,y,z$ is labeled as $\langle ij \rangle^\mu$ (see Fig.~\ref{fig:kitaev}). For each hexagonal plaquette, the local flux operator $W_p = \sigma_1^x \sigma_2^z \sigma_3^y \sigma_4^x \sigma_5^z \sigma_6^y$ is conserved by $H_K$, and has eigenvalues $\pm 1$ (zero or $\pi$-flux). Kitaev's solution involves writing each spin operator as the product of Majorana operators as follows: 
\beq \sigma^\mu_i = i b^\mu_i c_i, \text{ where } b^x_i b^y_i b^z_i c_i = 1 
\eeq 
in the physical subspace. The bond operators defined by $u_{\langle ij \rangle^\mu} = i b_i^\mu b_j^\mu$ have eigenvalues $\pm 1$, and commute with $H_K$. Each $u_{\langle ij \rangle^\mu}$ can be thought of as a $\mathbb{Z}_2$ valued lattice gauge field which couples the gauge-charged Majoranas. A theorem by Lieb\cite{Lieb} guarantees that the ground state of this model is in the flux-free sector ($W_p = 1$ for all plaquettes), where one can make a gauge choice of $u_{\langle ij \rangle^\mu} = 1$ ($i$ and $j$ belong to even and odd sublattices respectively). This leads to the free Majorana hopping Hamiltonian description for the bare Kitaev model, ${\cal H}_{\rm m}\approx{\cal H}_{\rm K}$, given by:
\beq
H_{\rm K} = -\frac{i J_K}{2} \sum_{j,\delta} c_j c_{j+\delta}.
\eeq
Here the sum is over all honeycomb sites $j$ and three nearest neighbors  $\delta$ shown in Fig.~\ref{fig:kitaev}. The single-particle excitation spectrum is given by $\varepsilon(\k) = J_K|1 + e^{i \k \cdot \n_1} + e^{i \k \cdot \n_2}|$, where $\n_1$ and $\n_2$ are basis vectors corresponding to the underlying Bravais lattice (see Fig.~\ref{fig:kitaev}). There are two Majorana cones at $K$ and $K^\prime$ points at the inequivalent corners of the Brillouin zone, which can be conveniently combined into a single Dirac cone at $K$. Expanding about the $K$ point in terms of continuum Dirac fields $\psi_{A/B}(\r)$, where $A$ and $B$ refer to the sublattice indices, 
\beq
c_i = \begin{cases}
\psi_A(\r) e^{i \bm{K} \cdot \r_i} + \mbox{h.c.}, ~ i \in A \\
\psi_B(\r) e^{i \bm{K} \cdot \r_i} + \mbox{h.c.}, ~ i \in B
\end{cases}
\eeq
and carrying out a gradient expansion of $H_K$, it is found that 
\beq
H_K = \sum_{\k} \psi_{\k}^\dagger (v\, \bm{\sigma} \cdot \k)  \psi_\k,
\eeq
where $v = 3 J_K / 2$, and $\psi_\k = (\psi_A(\k), \psi_B(\k))^T$. Thus, below the flux gap, the low-energy excitations of the bare Kitaev model are Dirac fermions. One can show that a uniform magnetic field $B \hat{z}$ acts as a Dirac mass for the fermions,\cite{Kitaev06} while a staggered magnetic field (switching signs between sublattices A and B) acts as a chemical potential.\cite{WCMB17} Hence, the Hamiltonian on adding a magnetic field that also breaks sublattice symmetry (or more generally, a time-reversal symmetry breaking term) is:
\beq
H_K = \sum_{\k} \psi^\dagger_{\k} (v\, \bm{\sigma} \cdot \k + m \sigma^z + \mu \sigma^0) \psi_\k, 
\eeq
with $m=0=\mu$ if TRS is present. For the solvable model, one needs the magnetic field to couple to all three spin components, and $m$ and $\mu$ are cubic on the applied field. However, on adding perturbations $J_H$ and $J_\Gamma$, this will no longer be necessary.\cite{SYB2016} Additionally, $m$ and $\mu$ are linear in the external magnetic field and exist for any field orientation. For $m > \mu$, we get a Chern insulator of spinons, while for $m < \mu$ we have a spinon Fermi surface. 

In order to find spin-spin correlators, one needs to find a representation of the spin-operators $\sigma^\mu$ in terms of the low-energy Majorana fermions. For the bare Kitaev Hamiltonian, this involves additional gapped flux excitations. However, additional perturbations, corresponding to the $J_H$ and $J_\Gamma$ terms in Eq.~(\ref{eq:kitaevhgamma}), renormalize the spin operator\cite{SYB2016} and result in an effective spin-operator of the form:
\beq
\sigma^a = \psi^\dagger m^a \psi  + \ldots.
\eeq
Here $m^a$ are $2 \times 2$ matrices whose specific form depends on the precise microscopic Hamiltonian, and the ellipsis correspond to gradient terms of $\psi$ which will give subdominant contributions to the spin-spin correlations. Since we are interested only in the scaling of the correlations with distance/frequency/temperature, and because $1/T_1$ has contributions from all $\langle \sigma^+\sigma^-\rangle$, $\langle \sigma^-\sigma^+\rangle$ and $\langle \sigma^z\sigma^z\rangle$, here we simplify the discussion by choosing 
$m^a = \sigma^0$. This reduces the problem to calculating density density correlations for nearly free Dirac fermions (the gapped gauge field can only mediate short-range interactions), for which analytical results are readily available.\cite{AVAPM2009} Choosing a different Pauli matrix $\sigma^a$, or treating the sublattice index more carefully, are expected to lead to the same scaling form of the retarded spin-spin correlations. 
Further, we always assume that all relevant energy scales ($T, \omega$) are much smaller than the gap $\Delta_v$ associated with 
$\mathbb{Z}_2$ magnetic flux excitations, also known as  
visons (defined by $W_p = -1$ in Fig. \ref{fig:kitaev}). Beyond these regimes, vison fluctuations can significantly affect the spin correlations, and the problem needs to be investigate numerically.\cite{KBM2018}

\begin{figure}
  \centering\includegraphics[scale=1.2]{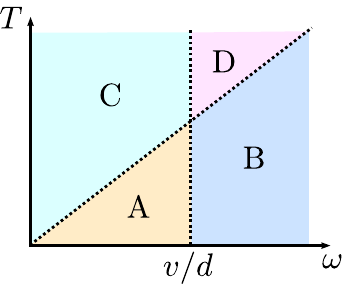}
  \caption{Schematic diagram indicating different regimes for $1/T_1$ as a function of $d$, $T$ and $\omega$ valid for Dirac spinons, see Eqs.(\ref{eq:dirac1}) and (\ref{eq:dirac2}), and spinon Fermi surfaces, see Eqs.(\ref{eq:fs1}) and (\ref{eq:fs2}). 
}
  \label{fig:regions}
\end{figure}

With all these considerations in mind, the low-energy dynamic spin-spin correlator  
in the $T \ll \omega$ limit is given by (see details in Appendix\,B): 
\beq
\mathcal{C}^{~''}_{+-}(\q,\omega) \approx 
\theta(\omega - v q) \frac{q^2}{\sqrt{\omega^2 - (v q)^2}}.
\label{eq:SCleanDirac}
\eeq
For calculating the relaxation time of a clean $\mathbb{Z}_2$ spin liquid with Dirac spinons, it is sufficient to use only ${\cal C}_{+-}$, as ${\cal C}_{-+}$ and ${\cal C}_{zz}$ will scale identically because of spin-rotation invariance of the spin liquid phase. As illustrated in Fig.\ref{fig:regions}, the emergent temperature scale set by the inverse distance, $\T_d = \hbar v /k_B d$, will be important for the discussion. Taking $v \approx 10^5$ m\,s$^{-1}$ as a typical velocity scale and $d$ between $1$ nm and 1 $\mu$m, $\T_d$ varies between $1-10^3$ K, which implies that both large and small $T/\T_d$ limits are experimentally accessible.  Considering different limits of $\omega/\T_d$ leads to the scaling 
\beq
\frac{1}{T_1} & \propto 
\begin{cases}
\omega^5 , ~ d \omega/v \ll 1, \\
\frac{1}{\omega d^6}, ~  d \omega/v  \gg 1, 
\end{cases}
\begin{array}{c} ({\rm A}) \\ ({\rm B}) \end{array}
\label{eq:dirac1}
\eeq
valid in the regime $\omega \ll T$ (regions A and B are shown in Fig.\ref{fig:regions}). We can also compute the relaxation 
time at large temperatures, but 
still smaller than the vison gap, i.e, $\omega \ll T \ll \Delta_v$, so that the contribution to spin correlations comes primarily from the nearly free gapless spinons. This leads to the scaling 
\beq
\frac{1}{T_1} & \propto    
\begin{cases}
\frac{T^2}{d^3}, ~  d \omega/v  \ll 1, \\
\frac{T^2 \omega^{5/2}}{\sqrt{d}} e^{-2\omega d/v}, ~  d \omega/v  \gg 1,
\end{cases}
\begin{array}{c} ({\rm C}) \\ ({\rm D}) \end{array}
\label{eq:dirac2}
\eeq
where regions C and D are shown in Fig.\ref{fig:regions}.

\tocless\subsubsection{Clean $\mathbb{Z}_2$ QSL with spinon Fermi surface}

A spinon Fermi surface can be obtained from the extended Kitaev honeycomb model by adding either a staggered magnetic field or three-spin interaction terms that breaks time-reversal symmetry (a uniform magnetic field results in a Chern insulator in the bare Kitaev model).\cite{Kitaev06,WCMB17} This implies that one takes the Hamiltonian in Eq.~(\ref{eq:pH}) with translation invariant $J_K$, $J_H$ and $J_\Gamma$ and $\bm{B}_i = \eta_i B \hat{z}$, where $\eta_i = \pm 1$ on the two sublattices of the honeycomb lattice. On different lattices, one can obtain a spin liquid with a Fermi surface even in presence of time-reversal,\cite{Hermanns2014,OBrien2016} and our results should hold irrespective of the physical origin of the spinon Fermi surface. One signature that one might expect from the spinon Fermi surface would be coming from the non-analyticity of the Lindhard function at momentum $q = 2 k_F$. This is different for Dirac fermions at finite chemical potential, as opposed to Fermi liquids with quadratic dispersion as emphasized in Refs.~\onlinecite{HDS2007,Wunsch2006}. However, for $k_F$ of the order of inverse lattice spacing, the signal is very small at accessible distances $d \gg 1/k_{F}$, and the main contribution again comes from the long-wavelength modes near $\q = 0$. Therefore, we do not need to consider both cases separately. To make analytical progress, we will assume that the chemical potential $\mu$ is the second largest energy scale in the problem after the vison gap ($\Delta_v \gg \mu \gg T, \omega, \T_d$). In the zero temperature ($T \ll \omega$) limit, this gives rise to a spin-spin correlation
\beq
\mathcal{C}^{~''}_{+-}(\q,\omega) \approx 
\theta(v q - \omega) \frac{q^2}{\sqrt{v^2 q^2 - \omega^2}} \frac{\omega \mu^2}{(vq)^3}.
\eeq
In the limit $T \ll \omega \ll \mu$, this results in a relaxation time given by:
\beq
\begin{array}{rl}
\displaystyle\frac{1}{T_1} \propto 
& \displaystyle 
\frac{\mu^2\omega^2}{2d}\left[ \frac{2d\omega}{v} K_0(2d \omega/v) + K_1(2d\omega/v)\right]  \\
\approx & \displaystyle   \begin{cases}
\frac{\omega}{d^2} , ~ d \omega/v \ll 1, \\
\frac{\omega^{5/2}}{\sqrt{d}} e^{-2\omega d/v}, ~  d \omega/v  \gg 1.
\end{cases}
\begin{array}{c} ({\rm A}) \\ ({\rm B}) \end{array}
\end{array}
\label{eq:fs1}
\eeq

In the high temperature ($\omega \ll T$) limit, the spin correlation takes the 
form:
\beq
\begin{array}{rl}
\mathcal{C}^{~''}_{+-}(\q,\omega) \approx & 
\displaystyle\Theta(vq - \omega) \, \frac{q^2}{\sqrt{v^2 q^2 - \omega^2}} \left[ \frac{2 \omega}{vq} K_1\left( \frac{vq}{2T} \right) \right] \\ 
& \displaystyle + \Theta(\omega - vq) \, \frac{q^2}{\sqrt{\omega^2 - v^2 q^2}} \frac{\pi}{2} \left[  1- e^{-\omega/2T}  \right].
\end{array}
\eeq
The relaxation time in this limit, assuming $\T_d \ll T$, is given by
\beq
\begin{array}{rl}
\displaystyle\frac{1}{T_1} \propto & \displaystyle \frac{\mu^2\omega^2}{2d}\left[ \frac{2d\omega}{v} K_1(2d \omega/v) + K_2(2d\omega/v)\right] \\
\approx & \displaystyle \begin{cases}
\frac{1}{d^3} , ~ d \omega/v \ll 1, \\
\frac{\omega^{5/2}}{\sqrt{d}} e^{-2\omega d/v}, ~  d \omega/v  \gg 1.
\end{cases}
\begin{array}{c} ({\rm C}) \\ ({\rm D})\end{array}
\end{array}
\label{eq:fs2}
\eeq
In the physically relevant regime $\omega \ll T \ll \mu$ and $ d \omega/v \ll 1$, the inverse relaxation time scales as $d^{-3}$ as a function of distance, and is independent of the frequency $\omega$ and temperature $T$ for the QSL with a spinon Fermi surface.

\begin{table}
\centering
\begin{tabular}{|c|c|c|c|c|}
\hline
    $\omega \ll T$      & \multicolumn{2}{c|}{$T$ dependence} & \multicolumn{2}{c|}{$d$ dependence} \\ \hline
          & Clean       & Dirty    & Clean         & Dirty   \\ \hline
$\mathbb{Z}_2$ Dirac  & $T^2$           & $T^{2-\alpha_1}$           & $d^{-3}$       & $d^{-3+\alpha_2}$     \\ \hline
$\mathbb{Z}_2$ FS &      $T^0$       & $T$           &      $d^{-3}$          & $d^{-2}$    \\ \hline
U(1) FS    & $T$     &     $T$     & $d^{-3}$     & $d^{-2}$ \\ \hline
\end{tabular}
\caption{Characteristic dependence of $1/T_1$ on the probe frequency $\omega$ and the sample probe distance $d$, considered in the limit $\omega \ll T$, with $d \omega/v \ll 1,~ \omega d^2 \ll D_s$ (see definition of $D_{s}$ in main text). $\alpha_i$ are positive numbers proportional to the disorder strength for weak disorder.}
\label{tab:gapless1}
\end{table}

\begin{table}
\centering
\begin{tabular}{|c|c|c|c|c|}
\hline
    $T \ll \omega$      & \multicolumn{2}{c|}{$\omega$ dependence} & \multicolumn{2}{c|}{$d$ dependence} \\ \hline
          & Clean         & Dirty    & Clean         & Dirty    \\ \hline
$\mathbb{Z}_2$ Dirac  & $\omega^5$           & $\omega^{5-\alpha_3}$           & $d^0$       &  $d^{-\alpha_4} $         \\ \hline
$\mathbb{Z}_2$ FS &      $\omega$       & $\omega$           &      $d^{-2}$          & $d^{-2}$         \\ \hline
U(1) FS    & $\omega$     &     $\omega$     & $d^{-3}$     & $d^{-2}$         \\ \hline
\end{tabular}
\caption{Characteristic dependence of $1/T_1$ on the probe frequency $\omega$ and the sample probe distance $d$, considered in the limit $T \ll \omega$ (the other limits/notations are identical to Table \ref{tab:gapless1}).}
\label{tab:gapless2}
\end{table}

\vspace{0.5 cm}
\tocless\subsubsection{Dirty $\mathbb{Z}_2$ QSL with Dirac spinons}

In the presence of quenched disorder, there is a drastic change in the nature of spin fluctuations in the extended Kitaev model. The precise characteristic of the change depends on the type of disorder introduced. A random bond disorder, corresponding to a random $J_K$ or $J_H$ to the Hamiltonian in Eq.~(\ref{eq:pH}), preserves time-reversal symmetry and translates to a random vector potential on the Dirac fermion,\cite{Willians_2010,WCMB17} while disorder that breaks time-reversal can induce either a random mass term or a random potential term. A slowly varying random mass term will result in energy gaps in most parts of the system, with gapless edge modes along the boundaries of the islands where the mass changes sign. These edge channels can be modeled as spinful charge-neutral Luttinger liquids, which will have their own signatures as discussed in Ref.~\onlinecite{Joaquin18}. On the contrary, a potential disorder will induce a lifetime for the Dirac fermions, and we will discuss this case in greater detail. 

First, we consider the case of time-reversal symmetric disorder. In this case, the disorder takes the form of a vector potential ($\bm{A}$) in the low-energy Hamiltonian. We further assume that the quenched vector potential disorder is short-range (delta-function) correlated in real space:
\beq
\begin{array}{c}
\displaystyle H = \sum_{\k, \k^\prime} \psi^\dagger_{\k} \left( v \, \bm{\sigma} \cdot \k \, \delta_{\k,\k^\prime} + \bm{A}_{\k - \k^\prime} \cdot \bm\sigma \right) \psi_{\k^\prime}, \\
\displaystyle \langle \bm{A}_{\q} \cdot \bm{A}_{\q^\prime} \rangle = (2\pi)^2 \delta(\q + \q^\prime) \Delta_A.
\end{array}
\eeq
The low energy behavior of Dirac fermions in the presence of vector potential disorder has been investigated in detail in Ref.~\onlinecite{Ludwig}. The system is described by a line of fixed points, characterized by scaling exponents that vary continuously with disorder. In particular, the dynamic critical exponent is found to be $\tz = 1 + \Delta_A/\pi$, and this difference in scaling of space vs time shows up in the relaxation time (note that a Dirac cone with no disorder has $\tz=1$). In the physically relevant regime of $\omega \ll T$, we can express the relaxation time as a scaling function, 
\beq
\frac{1}{T_1} \xrightarrow{\omega \ll T} T^{(6-\tz)/ \tz} \, \Psi_1(d \, T^{1/\tz}),
\eeq
where $\Psi_1(d T^{1/\tz})$
reflects the anomalous scaling of the relaxation time as a function of temperature and distance from the sample. An exact analytical expression for $\Psi_1$ in only available in the clean limit (see Appendix \ref{app:integrals2}). However, we note that the relaxation time will scale with some non-universal exponent of the distance that changes with disorder strength. At a fixed disorder strength, the noise data measured by changing 
$T$ and 
$d$ can be collapsed onto a single curve and tell us the value of the dynamic critical exponent $\tz$. 
 We add that an analogous calculation also gives us the scaling of the relaxation time as a function of $\omega$ in the zero temperature limit ($T \ll \omega$):
\beq
\frac{1}{T_1} \xrightarrow{T \ll \omega} \omega^{(6-\tz)/ \tz} \, \Psi_2(d \, \omega^{1/\tz}).
\eeq
We checked that for $\tz = 1$ (or, equivalently, $\Delta_A = 0$), corresponding to clean Dirac fermions, the scaling functions we analytically obtain give relaxation times that precisely match our previous results in the zero disorder case. For $\omega \ll T$, $\Psi_1(y) \sim y^{-3}$ in the clean limit. On adding disorder, we expect a correction to the power law which is proportional to the disorder strength, i.e, $\Psi_1(y) \sim y^{-3 - c_\Delta \Delta_A}$ for some constant $c_\Delta$ which, for weak disorder, 
is independent of the disorder strength. This insight can be used to work out the relaxation time for the dirty Dirac $\mathbb{Z}_2$ QSL to linear order in disorder strength, and we find that the following scaling holds:
\beq
\frac{1}{T_1} \propto \frac{T^{2 - (4 - c_\Delta) \Delta_A}}{d^{3 - c_\Delta \Delta_A}}, ~ \omega d/v \ll 1.
\eeq
Some intuition for these results can be obtained by studying the increase in the density of states per unit area $\rho(\omega)$ at low energy. In presence of disorder, a simple scaling argument shows that $\rho(\omega) \propto \omega^{(2-\tz)/\tz}$ where the exponent is less than one for $\tz > 1$.\cite{Ludwig} Hence, we expect the dirty Dirac $\mathbb{Z}_2$ QSL interpolates between the clean Dirac $\mathbb{Z}_2$ QSL and the $\mathbb{Z}_2$ QSL with a spinon Fermi surface (which we study next).  
Analogous arguments can be used for the scaling function $\Psi_2$ to find the temperature and distance scaling of the relaxation time in the $T \ll \omega$ limit --- the results are presented in Table \ref{tab:gapless2}. 

If the disorder breaks time-reversal symmetry, then both potential and mass disorder are allowed along with the vector potential disorder. A detailed discussion on the effects of these different kinds of disorders are contained in Ref.~\onlinecite{EversMirlin_RMP}. For instance,
random mass disorder 
turns out to be marginally irrelevant and the system flows back to the clean limit. This would be the case for random (but unidirectional) magnetic fields in the Kitaev model. In the case of random potential disorder, the system is in the Wigner-Dyson symplectic class with an additional topological $\theta$ term that leads to delocalization. In this limit, a disordered Dirac fermion system is argued to behave like a metal. In the absence of precise analytical results for the susceptibility, we conjecture that the spin correlations would exhibit a diffusive behavior. Therefore, the relaxation time should have the same behavior as the dirty spinon Fermi surface case discussed next. At least two of these types of disorder automatically generates the third type by a renormalization group flow,\cite{EversMirlin_RMP} and the system flows to the IQH transition fixed point. The lack of analytical knowledge of the critical exponents at this fixed point renders it difficult to make a prediction for the scalings of the relaxation time in this limit.

\vspace{0.5 cm}
\tocless\subsubsection{Dirty $\mathbb{Z}_2$ QSL with spinon Fermi surface}
Finally, we consider the case of a $\mathbb{Z}_2$ spin liquid with a spinon Fermi surface in the presence of weak disorder. This can be realized within the Kitaev honeycomb model with a staggered magnetic field, which induces a Fermi surface as discussed before. In the presence of disorder and short range interactions mediated by the gapped $\mathbb{Z}_2$ gauge field, the spin susceptibility at low $T$ takes the following diffusive form\cite{Castellani1986} that holds as long as the relevant energy scales $T, \omega$ are much smaller than the Fermi energy. 
\beq
\mathcal{C}^{~''}_{+-}(\q,\omega) \approx - \text{Im}\left[ \frac{\nu D_s q^2}{-i \omega + D_s q^2} \right] = \frac{\nu D_s q^2 \omega}{\omega^2 + D_s^2 q^4},
\eeq
where $\nu$ is the spinon density of states at the Fermi surface, and $D_s$ is the spin-diffusion constant. Using this form, we can again calculate the relaxation time in the limits of large and small distance $d$. In the zero temperature limit ($\omega \gg T$), $1/T_1$ takes the form:
\beq
\frac{1}{T_1}& \approx &  \begin{cases}
\frac{\omega}{d^2} , ~ \omega d^2 \ll D_s \\
\frac{1}{\omega d^6}, ~  \omega d^2 \gg D_s
\end{cases}
\eeq
For $T \ll \omega \ll \mu$, this behaves as:
\beq
\frac{1}{T_1}& \approx &  \begin{cases}
\frac{T}{d^2} , ~ \omega d^2 \ll D_s \\
\frac{T}{\omega^2 d^6}, ~  \omega d^2 \gg D_s
\end{cases}
\eeq
In a typical dirty spin liquid with a spinon Fermi surface, we expect that $D_s = v_F \ell/2 \approx 10^{-4}$ m$^2$ s$^{-1}$, assuming a spinon Fermi velocity of $10^5 $m s$^{-1}$ and a mean-free path $\ell$ of tens of lattice spacing. On the other hand, for existing spin qubits $\omega \approx 10^9 $Hz and a sample-probe distance which can vary between from nanometers to microns, an upper bound of $10^{-5}$ m$^2$ s$^{-1}$ can be found for $\omega d^2$. Therefore, $\omega d^2 \ll D_s$ is the physically relevant limit for measurements.

\vspace{0.5 cm}
\tocless\subsubsection{Clean U(1) QSL with spinon Fermi surface}
Certain quantum spin liquids are described by Dirac cones or Fermi surfaces of low energy fermionic spin-half spinons with a conserved spinon number. These phases have gapless neutral spin-carrying fermionic excitations at a Fermi surface or Dirac cones, and these spinons are strongly coupled to an emergent dynamical compact U(1) gauge field. Hence, these phases have U(1) topological order, with gapless photons and a novel gapped magnetic monopole.\cite{SavaryBalentsQSLReview,LeeNagaosa} At temperature and energy scales far below the monopole gap, the spin dynamics are controlled by the gapless spinons which are strongly renormalized by gauge-field fluctuations. 

Here, we focus on a U(1) QSL with a spinon Fermi surface. Such a phase is described by an action of quantum electrodynamics in two spatial and one time dimensions (QED$_3$): 
\beq
S &=& \int dt d\r \bigg[ f^\dagger_{\r,\sigma} (\partial_t - i a_{0} - \mu) f_{\r,\sigma} \nn && ~~~~~~~~~ + \frac{1}{2m} f^\dagger_{\r,\sigma} (\mathbf{\nabla} - i \mathbf{a})^2 f_{\r,\sigma}  + \frac{1}{4g^2} (\epsilon_{\mu \nu \lambda} \partial_{\nu} a_{\lambda})^2 \bigg], \nn
\eeq
where the low energy $f$ fermions (with effective mass $m$) are related to the spin operator $\bS(\r)$ by $\bS(\r) = f^\dagger_{\r,\alpha} \mathbf{\sigma}_{\alpha \beta} f_{\r,\beta}$, $\mu$ is the chemical potential, $m$ is an effective mass and $a_{\mu}$ is an emergent U(1) gauge field, with $\epsilon_{\mu \nu \lambda} \partial_{\nu} a_{\lambda}$ being the corresponding field strength tensor.

This problem has been studied extensively, and the clean system can be described by a strong coupling fixed point.\cite{Polchinski1994} The large N expansion which can justify RPA has been shown to be uncontrolled.\cite{SSLee} However, the higher loop corrections which also contribute to the same order should leave the relative scaling of momentum, frequency and temperature unchanged.\cite{SSLee} Therefore, using the RPA results should not affect the scaling of the relaxation time although it may affect the exact numerical prefactors. With this prelude, we use the RPA spinon Green's function given in Ref.~\onlinecite{LeeNagaosa}, assuming a quadratic dispersion $\xi_{\k} = \frac{k^2}{2m} - \mu$ for the fermionic spinons:
\beq
G_f(\k, \omega) &=& \frac{1}{\omega - \xi_{\k} - \Sigma(\omega)}, \\
\text{ where } \Sigma^{~''}(\omega) &=& - C \omega^{2/3} \text{ for } \omega > 0, \text{ and } ~ C \approx \mu^{1/3}. \nn
\eeq
The dominant contribution to the dynamic spin susceptibility at low energies can be computed from four point functions of the $f$ fermions, comes from the anomalous imaginary part of the self-energy due to scattering by fluctuating gapless gauge bosons. The spin-spin correlations at low energy in a U(1) QSL is given by (see Appendix \ref{app:Integrals})
\beq
\mathcal{C}^{~''}_{+-}(\q,\omega) \approx \frac{\omega}{\sqrt{v_F^2 q^2 + C^2 \varepsilon^{4/3}}}, 
\eeq
where $\varepsilon = \text{max}(\omega,T)$. 
The relaxation time is given in different limits by the following expressions when $\omega \gg T$:
\beq
\frac{1}{T_1}& \approx &  \begin{cases}
\frac{\omega}{d^3}, ~  \omega d/v_F \ll \left(\frac{\omega}{\mu}\right)^{1/3} \ll 1, \\
\frac{\omega^{1/3}}{d^4} , ~ \omega d/v_F \gg 1.
\end{cases}
\eeq
In the alternate limit of $\omega \ll T$, the correlation function is dominated by $T$ and hence the relaxation times are as follows:
\beq
\frac{1}{T_1}& \approx &  \begin{cases}
\frac{T}{d^3}, ~  \omega d/v_F \ll T d/v_F \ll \left(\frac{T}{\mu}\right)^{1/3} \ll 1 \\
\frac{T^{1/3}}{d^4} , ~ T d/v_F \gg 1
\end{cases}
\eeq

\tocless\subsubsection{Dirty U(1) QSL with spinon Fermi surface}
Introduction of disorder to the clean U(1) spin liquid with a Fermi surface of spinons is likely to lead to a flow away from the $z = 3$ critical point to a diffusive Fermi liquid of spinons.\cite{Galitski} Since the gauge field does not contribute directly to spin susceptibility except via its effect on renormalization of the spinon energy, we expect an identical behavior for the relaxation time as the $\mathbb{Z}_2$ spin liquid with disorder as discussed earlier.

\vspace{1 cm}

\subsection{Disordered VBS states}
Frustrated magnets which do not order at low temperatures can also have ground states that spontaneously break translation symmetry (and possibly certain rotation symmetries) of the lattice, but preserve spin-rotation symmetry. These paramagnetic states are called valence bond crystals or valence bonds solids (VBS).\cite{SB1990,NRSS90} Our focus in this section is on gapless systems, which is the likely fate of VBS states in presence of disorder.\cite{Itamar1,Itamar2} However, we first study a clean VBS phase to set the stage. Such phases typically have gapped particle-like triplon excitations with quadratic dispersion near the band minimum with some effective mass $m$.\cite{SB1990,Uhrig2004} The retarded spin-spin correlator at low energy and temperatures ($\omega,T \ll J$, where $J$ is the spin-exchange scale) is dominated by single bosonic triplon excitations:
\beq
\mathcal{C}^{~''}_{+-}(\q,\omega) = \delta(\omega - \Delta_T - q^2/2m).
\eeq
This implies that the relaxation rate detected by the spin probe 
can be calculated to be:
\beq
\frac{1}{T_1} \approx m^2(\omega - \Delta_T) e^{-2d \sqrt{2m(\omega -\Delta_T)}}   \Theta(\omega - \Delta_T) 
\eeq
This relaxation rate is similar to a trivial thermal paramagnetic phase: it is non-zero above a threshold frequency equal to the spin gap $\Delta_T$, and shows exponential decay with sample-probe distance $d$. Further, there is a distinct dip in $T_1$ at an optimal distance of $d = [2m(\omega - \Delta_T)]^{-1/2}$ which can be used to estimate the the effective mass of the low-energy triplons. 

\begin{figure}[]
    \centering
    \includegraphics[scale=0.6]{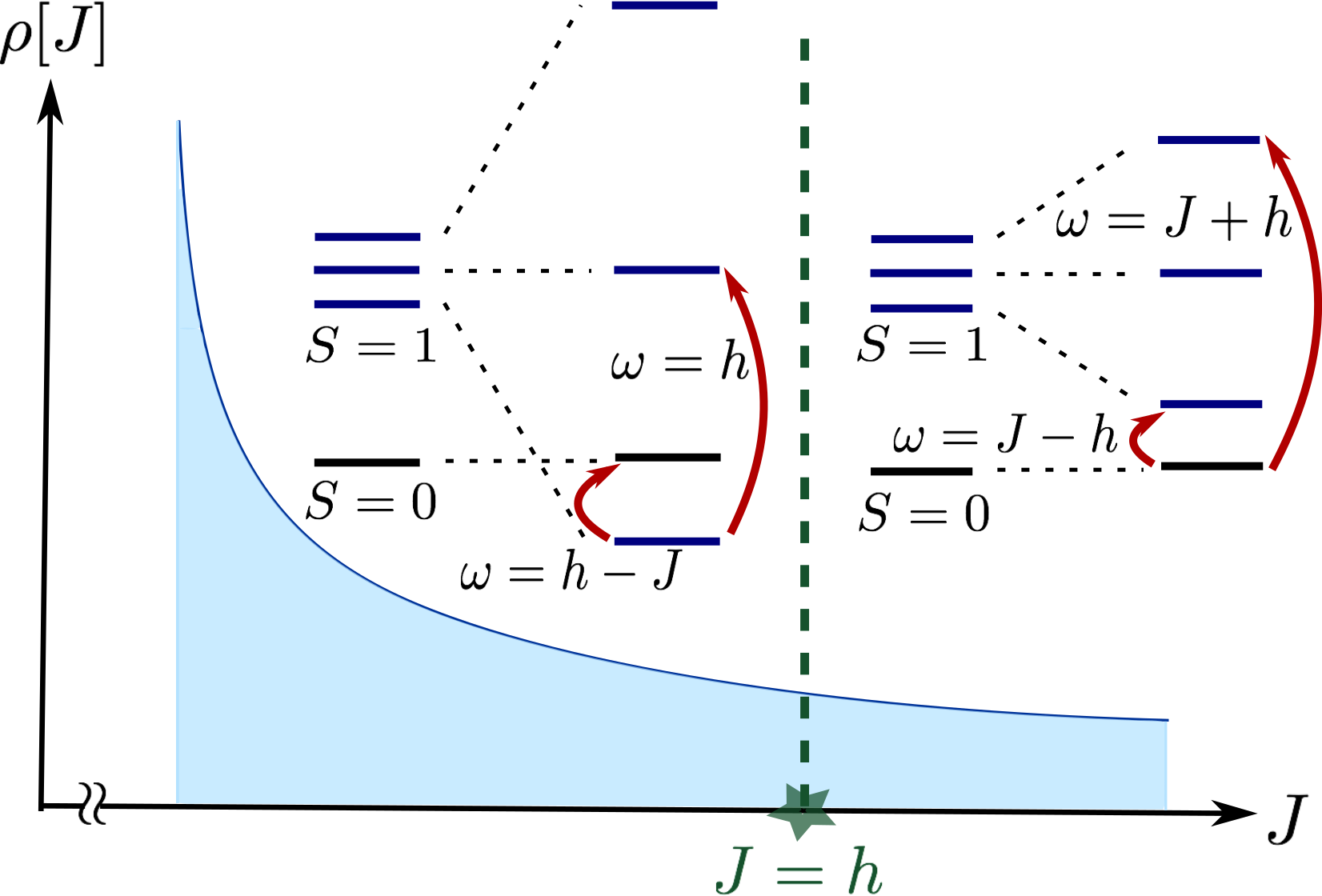}
    \caption{The power law distribution of couplings $J$, with the level structures of a pair of weakly coupled impurity spins at $J>h$ and $J<h$ showing the transitions that contribute to the dynamic spin-structure factor.}
    \label{fig:DirtyVBS}
\end{figure}

Certain materials have the additional complication of quenched randomness, which can give rise to random strengths of magnetic exchanges. A theory for such disordered frustrated quantum magnets was provided in Ref.~\onlinecite{Itamar1}. In the presence of weak random bond disorder, a gapped quantum spin liquid state is typically stable. However, a paramagnetic valence bond solid crystal is unstable to nucleation of vortices in the VBS order parameter. For topological reasons, each such defect will carry a dangling spin-half. In the opposite limit of strong disorder, the naive expectation of a paramagnetic phase made of randomly pinned singlets (called a 'valence bond glass' in Ref.~\onlinecite{Itamar1}) fails, and spinful defects are nucleated. Thus, Ref.~\onlinecite{Itamar1} argued that in two very different limits (and hence possibly also at intermediate disorder), the system may be described at low energy scales by a random network of defect spins with a broad distribution of exchange coupling.

 This small subsystem of defect spin-half moments dominate the thermal and quantum fluctuations at low-temperatures. This leads to several interesting observable consequences, the most prominent being the power law behavior of specific heat as a function of temperature with a sub-linear non-universal exponent. Further, Ref.~\onlinecite{Itamar2} discussed the effect of Dzyaloshinskii-Moriya (DM) interactions on these systems, which can modify the scaling of specific heat and other observable properties. Below, we calculate the contribution to magnetic noise of these defect spins will be the dominant source of noise at any frequency $\omega \ll \Delta_S$, where $\Delta_S$ is the gap of the clean VBS phase.

The low energy dynamics of the system, in absence of spin-orbit coupling, is described by a random-bond Heisenberg model. This can be obtained by switching off all terms except the Heisenberg exchange in our parent Hamiltonian in Eq.~(\ref{eq:pH}), i.e, by setting $J_K = J_\Gamma = 0 = \bm{B}_i$:
\beq
H_{H} = \sum_{i,j} J_{ij} \mathbf{S}_i \cdot \mathbf{S}_j, ~~ J_{ij} = \bar{J} + \Delta J_{ij}
\eeq
The typical distance between the defect spins is given by the correlation length of the VBS order parameter, which is given by $\xi/a \sim \exp[C_\xi J^2/\langle \Delta J^2\rangle]$,\cite{Binder1983} where $a$ is some appropriate microscopic lengthscale like the lattice constant, and $C_\xi$ is a numerical constant of order unity. The physics at lengthscales greater than $\xi$ is therefore described by the appropriate RG flow of this random bond Heisenberg model. While this is not well-controlled in two dimensions, Ref.~\onlinecite{Itamar1} argues that, for large lengthscales, we can treat this problem as spins interacting with a continuous distribution of couplings with a broad power-law tail that decays with a non-universal exponent.\cite{BhattLee1981,BhattLee1982} The zero-field specific heat of these materials over a broad range of temperatures shows a power law behavior $T^\alpha$, which is argued to arise from a density of couplings $\rho[J] = J^{\alpha - 1}$ ($\alpha < 1$) of the defect spins.\cite{Itamar1} Assuming that these spins are quite dense on the lattice scale (the VBS correlation length is small, as it seems to be for YbMgGaO$_4$), the spin qubit, placed sufficiently far away, will be able to sense fluctuations from the spins which are weakly coupled. Hence, although these defect spins are localized (not long-wavelength modes), yet useful information can be obtained by looking at the spin dynamics at long lengthscales as a function of an applied magnetic field.

Ref.~\onlinecite{Itamar1} gives the transverse part of the dynamic structure factor, $S_T[J,\mathbf{R}](\q, \omega) = S_{+-}[J,\mathbf{R}](\q,\omega) + S_{-+}[J,\mathbf{R}](\q, \omega)$, for two spins with an effective exchange $J$ at distance $R$, in presence of an applied field $B$ as follows (letting $h = g \mu_B B$):
\begin{widetext}
\beq
\begin{array}{rl}
\displaystyle S_T[J,\mathbf{R}](\q, \omega) =  & \displaystyle \Theta(J - h) \left( \frac{1 - \cos(\q \cdot \mathbf{R})}{2} \right) \sum_{\pm} \delta(\omega - J \pm h) \\
& \displaystyle + \Theta(h - J) \left[  \left( \frac{1 - \cos(\q \cdot \mathbf{R})}{2} \right) \delta(\omega + J - h)  + \left( \frac{1 + \cos(\q \cdot \mathbf{R})}{2} \right) \delta(\omega - h) \right].
\end{array}
\label{eq:SJR}
\eeq
\end{widetext}
The full transverse structure factor may then be obtained by integrating Eq.~(\ref{eq:SJR}) over the density of states $\rho[J] \sim J^{\alpha -1}$: 
\beq
S_T(\q, \omega) = \int_J \rho[J] \, S_T[J,\mathbf{R}[J]](\q, \omega) ,
\eeq
The function $\mathbf{R}[J]$, which quantifies the singlet size as a function of its energy splitting, is not exactly known for two-dimensional systems. However, we do not need exact knowledge of $\mathbf{R}[J]$ if we assume statistical rotation symmetry of our system, which is reasonable for random location of the defects. In this case, the angular integral $\cos(\q \cdot \mathbf{R})$ vanishes when integrated over $\q$ (as the prefactor depends only on $q$). We can see that the noise at low temperatures $T \ll h, \omega$ is given by:
\beq
\begin{array}{rl} 
\displaystyle\frac{1}{T_1} \approx & \displaystyle\frac{1}{d^4} \bigg[ \rho(\omega + h) + \theta(\omega - 2h)\rho(\omega - h) \\ 
& \displaystyle + \theta(h - \omega) \rho(h - \omega) + \frac{h}{\alpha}\rho(h) \delta(\omega - h) \bigg].
\end{array}
\eeq
The most noticeable feature is the delta function at the Zeeman energy, which is due to the resonance between $S_z = 0$ and $S_z = 1$ triplet states. The distribution of effective $J$ ensures that it will be present at any magnetic field, with a gradually increasing strength till the field hits a saturation value. The relaxation time will accordingly show a sharp drop at this resonance. The other noticeable features are the step-functions that arise from the distinct singlet-triplet transitions that contribute to the spin correlations for $J < h$ and $J > h$ (see Fig. \ref{fig:DirtyVBS}). For example, for $J < h$, the triplet to singlet transition has a frequency $\omega = h - J$, which always contributes because of the power-law distribution of $J$ across the system. Therefore, there is an associated step function $\Theta(h - \omega)$ coming from the positivity of $J$, and the corresponding density of states $\rho(J = h - \omega)$. Such distinctive step functions may also be accessed via tuning of the magnetic field from small to large values. 

For completeness, we also provide a computation of the longitudinal structure factor. We can again find it for the two-spin Hilbert space with separation $\mathbf{R}$ and effective coupling $J$, and then integrate it over the density of states:
\beq
\begin{array}{rl}
S_{zz}[J,\mathbf{R}](\q, \omega) = & \displaystyle\frac{1}{4}\big[ \theta(J - h)(1 - \cos(\q \cdot \mathbf{R})) \delta(\omega - h)  \\
& \displaystyle +  \theta(h - J) (1 + \cos(\q \cdot \mathbf{R})) \delta(\omega) \big],
\end{array}
\eeq
where
\beq
S_{zz}(\q, \omega) & = & \int_J \rho[J] S_{zz}[J,\mathbf{R}[J]](\q, \omega).
\eeq
Since the probe 
only detects at a finite frequency $\omega > 0$, the relevant contribution to the relaxation time is again the delta-function at $\omega = h$, which has the same physical effect on $T_1$ as the transverse correlators, namely, a sharp drop in $T_1$ at this resonance. 

We reiterate that our results are valid for the lowest temperature scales ($T \ll J$) when the random singlet phase is a good description of the system. At higher temperatures, we expect the spin-correlations to become unimportant, and the defect spins to behave as nearly free spins. In other words, the dimer physics is replaced by independent fluctuating spins, and the phase is no longer distinct from a thermal paramagnet. 

\section{Detection of anyonic statistics in gapped systems}
\label{sec:anyons}
In two spatial dimensions, the quantum mechanical wave-function of two identical particles can pick up phase factors different from $\pm 1$ when the particles are exchanged (or braided) adiabatically.\cite{Leinaas1977} These particles are said to have anyonic statistics. Each anyon can be thought of as a flux-charge composite, with a statistics parameter $\alpha$ which indicates that an adiabatic exchange of two identical gapped anyons results in a phase factor of $e^{i \pi \alpha}$ in the quantum wave-function of the state.\cite{Wilczek1,Wilczek2} In two-dimensions, $\alpha$ can be arbitrary, in contrast to three (or higher) dimensions where $\alpha$ is either zero (bosons) or one (fermions). Inspite of their discovery in quantum Hall states long ago, experiments to directly detect anyonic statistics are challenging. 
In this section, we argue that spin probes can detect anyonic statistics provided these anyons arise as emergent low energy spin carrying excitations in an insulator. This is true, for example, for the chiral spin liquid where the emergent quasipaticles have semionic statistics ($\alpha = 0.5$).\cite{LevinGu_2012,BMP14,QGY16} Such a phase has also been proposed within the framework of the Kitaev-Heisenberg-$\Gamma$ model that we discussed in Eq.~(\ref{eq:kitaevhgamma}), in presence of a magnetic field.\cite{LiuNormand2018} 

For phases with gapped spin excitations, we need to tune the energy-gap $\omega$ of the probe so that it is larger than the minimum gap to local excitations in order to have accessible relaxation rates. The relaxation rate as a function of the probe energy-gap at a fixed temperature $T$ provides crucial information about the statistics of the particles. In particular, the threshold behavior at the frequencies close to the spin gap at low temperature has a universal power law growth where the exponent is fixed by the braiding statistics of the anyonic excitations and is robust to short-range interactions. 

\subsection{Free bosons}
As a warm-up problem, we first discuss a simple phase where the emergent low energy degrees of freedom are gapped fractionalized spin-half excitations with bosonic self statistics. These excitations, called spinons, are characteristic of frustrated spin-models in phases that exhibit $\mathbb{Z}_2$ topological order.\cite{ReadSachdev} They are similar to the $e$ particle in the toric code,\cite{Kitaev03} but they also carry spin-half in addition. One can describe such a phase using the Schwinger boson representation of $S=1/2$ spins,\cite{AA1988,ReadSachdev,SSkagome} where the spin operator $\bS(\r)$ is written in terms of bosonic operators $b_{\r,\alpha}$ with an additional local constraint.
\beq
\bS(\r) = \frac{1}{2} b^\dagger_{\r,\alpha} \bm{\sigma}_{\alpha \beta} b_{\r,\beta}, \quad b^\dagger_{\r,\alpha}b_{\r,\alpha} = 1.
\eeq
Such models also have emergent fluxes on the underlying $\mathbb{Z}_2$ gauge field, analogous to the $m$ particle of the toric code. These fluxes, called visons, do not carry any spin, have bosonic self-statistics and have semionic mutual statistics with the spinons. In the regime where the vison gap $\Delta_v$ is much larger than the spinon gap $\Delta_s$ and the temperature $T$, we can safely neglect the visons as their only effect is to induce short-range interaction between the spinons. Although such interactions do affect bosons, as a first approximation we treat the spinons as approximately free quasiparticles. In this limit, we can write the spinon Green's function, defined by $G_s(\r,\tau) = -\langle T_\tau[b_{\r,\alpha}(\tau) b_{0,\alpha}^\dagger(0)] \rangle $, as that of a free boson, with a generic quadratic dispersion above the gap $\Delta_s$. Converting to momentum and Matsubara frequency, it takes the following familiar form close to the band minimum ($m$ is the effective mass):
\beq
G_s(\k, i \on) = \frac{1}{i \on - \xi_\k}, \text{ where } \xi_{\k} = \Delta_s + \frac{\k^2}{2m}.
\eeq
The retarded spin-spin correlator $C^R_{+-}(\q, \omega)$ can be found by analytic continuation of the Matsubara correlator $C_{+-}(\q, i\on)$:
\begin{widetext}
\beq
\mathcal{C}_{+-}(\q, i\on) & = & \frac{1}{\beta L^2} \sum_{\k, i \Omega_n} G_s(\k + \q, i \Omega_n + i \omega_n) G_s(-\k, - i \Omega_n) = \int \frac{d^2k}{(2\pi)^2} \frac{1 + n_B(\xi_\k) + n_B(\xi_{\k + \q})}{i \on - \xi_{\k} - \xi_{\k + \q}} \nn
& \xrightarrow{T \rightarrow 0} &  \int \frac{d^2k}{(2\pi)^2} \frac{1}{i \on - \xi_{\k} - \xi_{\k + \q}}  \nn
\implies \mathcal{C}^{~''}_{+-}(\q, \omega) & \xrightarrow{T \rightarrow 0} & - \text{Im}\left[ C_{+-}(\q, i\omega_n \rightarrow \omega + i0^+) \right] = \pi  \int \frac{d^2k}{(2\pi)^2} \, \delta(\omega - \xi_{\k} - \xi_{\k + \q})  = \frac{m}{4} \Theta\left( m( \omega - 2\Delta_s) - \frac{q^2}{4m} \right) \nn
\eeq
\end{widetext}
From this, the relaxation time can be calculated for $\omega \gg T$, when $ \coth(\omega/2T) \approx 1$. Leaving the detailed interpolating functions to Appendix \ref{app:Integrals}, here we focus on certain limits.
\beq
\frac{1}{T_1} & \propto & \begin{cases}
(\omega - 2\Delta_s)^2 \Theta\left( \omega - 2\Delta_s \right), ~ Qd \ll 1 \\
\frac{1}{d^4} \Theta\left( \omega - 2\Delta_s \right), ~ Qd \gg 1
\end{cases}
\eeq
Here $Q = \sqrt{4m(\omega - 2 \Delta_s)}$ denotes a momentum scale corresponding to excitation energy above the spin-gap of $2 \Delta_s$ and which limits the $q$ integral. Since our continuum approximation to the dispersion holds close to the bottom of the band, we expect the $ Qd \ll 1$ limit to be more accessible. In this limit, the relaxation-time is independent of the distance $d$ and grows as a power law with $\omega - 2 \Delta_s$, the energy above the threshold. 

\subsection{Anyons}
The non-local nature of the anyons implies that a single isolated anyon cannot be created locally. We assume that any local quantum fluctuation creates a couple of anyons, as in chiral QSLs. The braiding phase $\alpha$ that arises from the exchange of two anyons can be theoretically characterized by a Chern Simons vector potential $\bm{a} = (c \alpha/q) \nabla \phi$, where $\phi$ is the angle made by the vector connecting the two anyons (relative to an arbitrary reference), $q$ is their charge under the Chern Simons gauge field and $c$ is the speed of light. This term takes care of the exchange statistics, or in other words, mediates a long-range statistical interaction between the anyons while the Hamiltonian acts on bosonic wave-functions.\cite{Arovas1985} For a pair of anyons with quadractic dispersion, the Hamiltonian is:
\beq
H = \frac{\bm{P}^2_{\bm{R}}}{2m} + \frac{p^2_r}{m} + \frac{(p_{\phi} - \alpha)^2}{mr^2} + V(r,\phi)
\eeq 
where $\bm{R}$ is the center of mass coordinate, $\r = (r, \phi)$ is the relative coordinate, $m$ is the mass of each anyon and $V(r,\phi)$ represents some short range interaction between the anyons. Such a formulation was used by Ref.~\onlinecite{Sid_PRL2017} to write down a robust expression for the general two-anyon structure factor at the threshold of the gap. While such structure factors are accessible by neutron scattering in principle, the threshold behavior requires a probe with excellent energy resolution at low energies. Spin qubits are well-suited for this purpose, and hence we use the results for the correlation function of local bosonic operators (including the spin operator) in presence of a Chern-Simons field from Ref.~\onlinecite{Sid_PRL2017} to calculate the relaxation rate, again working under the assumption that $T$ is smaller than the spin-gap (which is $2\Delta_s$ in our convention):
\begin{widetext}
\beq
\mathcal{C}^{~''}_{+-}(\q,\omega) &\propto& J_{\alpha}^2(a \sqrt{m(\omega - 2\Delta_s) - q^2/4}) \, \Theta\left( m(\omega - 2\Delta_s) - q^2/4 \right), 
\eeq
which results in 
\beq
\frac{1}{T_1} & \propto & \begin{cases}
(\omega - 2\Delta_s)^{2+\alpha} \, \Theta\left( \omega - 2\Delta_s \right), ~ Qd \ll 1 \\
\frac{(\omega - 2\Delta_s)^\alpha}{d^4} \Theta\left( \omega - 2\Delta_s \right), ~ Qd \gg 1
\end{cases}
\eeq
\end{widetext}
In particular, we see that we recover our result for free bosons with $\alpha = 0$, where $1/T_{1}$ at the threshold $\omega \gtrsim 2\Delta_s$ is set by $(\omega - 2\Delta_s)^{2}$ for small $d$, and by $d^{-4}$ for large $d$. For general anyons with a statistics parameter $\alpha$, the power law at the threshold is modified to be $2 + \alpha$, which provides a striking signature for detecting anyonic statistics in gapped phases of 2d quantum matter. We specifically point out that for gapped fermions, $\alpha = 1$ and the relaxation time at the threshold is proportional to $(\omega - 2\Delta_s)^{3}$. Insulating phases with gapped charge-neutral fermionic excitations occur in the Kitaev honeycomb model in anisotropic limits or in presence of additional time reversal symmetry breaking terms, like a uniform magnetic field discussed in Eq.~(\ref{eq:pH}).\cite{Kitaev06,SYB2016}

The relaxation times shown a different power law dependence on quasiparticle statistics in the limit $Qd \gg 1$. However, the validity of our low energy expressions for the dynamic spin structure are doubtful in those limits, and a more elaborate computation of the relaxation rate is required to address this more accurately.
 
\subsection{Effects of interaction}

Short range interactions do not affect the structure factor for non-bosonic anyons at low energies, due to the rigidity of the two-anyon wavefunction at short distances where the interactions are the strongest, as argued in Ref.~\onlinecite{Sid_PRL2017}. This implies, for example, that in a chiral spin liquid phase with semionic low-energy excitations, we should still see an inverse relaxation time that goes as $(\omega - 2\Delta_s)^{5/2}$, even if we add in effects of interactions. However, free bosons are affected crucially by short-range replusive interactions as they lack any statistical replusion.\cite{Sid_PRL2017} As seen both numerically\cite{Sid_PRL2017} as well as in field-theoretic calculations,\cite{QXS09} for bosons with short range repulsive interactions  $\mathcal{C}^{~''}_{+-}(\q, \omega)$ receives a log squared correction:
\beq
\mathcal{C}^{~''}_{+-}(\q,\omega) \approx \frac{\Theta\left( \omega - 2\Delta_s - q^2/4 \right)}{\left[ \ln\left( [4m(\omega - 2\Delta_s) - q^2] b^2/16\right) + 2 \gamma \right]^2 + \pi^2}, \nn
\eeq
where $\gamma$ is the Euler Mascheroni constant and $b$ is an effective range of interaction. This leads to a correction in the relaxation time for weakly interacting gapped bosons, which can be analytically computed in the limit when $Qb \ll 1$ ($Q = \sqrt{4m(\omega - 2 \Delta_s)}$ is the typical momentum scale of excitations). Since the interaction range $b$ is of the order of a few lattice spacing ($b \approx a$) while the typical excitation wavelength $Q^{-1}$ needs to much larger than the lattice spacing/interaction range for the threshold behavior to hold, this assumption is well-justified for most short-range interactions:

\beq
\frac{1}{T_1} & \propto & \begin{cases}
\frac{(\omega - 2\Delta_s)^{2}}{\ln^2(Qb)} \, \Theta\left( \omega - 2\Delta_s \right), ~ Qd \ll 1 \\
\frac{1}{d^4 \, \ln^2(Qb)} \Theta\left( \omega - 2\Delta_s \right), ~ Qd \gg 1
\end{cases}
\eeq

If the anyons carry charge under an external or emergent gapless gauge field, then long range power law interactions (like Coulomb) can affect the relaxation time significantly. We will not solve the problem here in full generality, but we note that an analogous detection scheme for gapped magnetic monopoles interacting via gapless photons in spin-ice materials have been suggested in Ref.~\onlinecite{Kirschner2018}.

\section{Implications for material candidates}
\label{sec:materials}
In this section, we discuss the implications of our results for specific materials. Indeed, a large number of candidates exist for the different phases of quantum magnetism we have discussed so far, including the more exotic ones with topological order.\cite{Lee,Trebst2017}  

We chose to work on the perturbed Kitaev models, because spin-orbit coupled honeycomb lattice iridates provide an avenue to realizing such spin liquid states.\cite{Jackeli,Chaloupka,Trebst2017,KitaevSLReview} The Kitaev interaction is dominant in spin-orbit coupled iridates like $\alpha$-Na$_2$IrO$_3$,\cite{Chun2015} $\alpha$-Li$_2$IrO$_3$,\cite{Williams2016}  $\alpha$-RuCl$_3$.\cite{Plumb2014} None of these materials are exactly described by a bare Kitaev Hamiltonian, but the dominant Kitaev interaction is expected to lead to a stable spin-liquid phase where the exotic nature of excitations is independent of the details of the parent Hamiltonian. Indeed, neutron scattering and transport experiments see strong signatures of the spin liquid phase being present at large magnetic fields that suppresses  magnetic ordering in $\alpha$-RuCl$_3$.\cite{Nagler1,Nagler2,ThermalHall_RuCl3_Matsuda} Questions regarding the impact of phonons, however, remain a challenge. Theoretical proposals also predict the possibility of different spin liquid phases as a function of the magnetic field, including one with a spinon-Fermi surface \cite{ZouHe2018,Jiang2018,NPatel2018}. A noise magnetometry study, which depends solely on the spin sector, would help to isolate the true nature of the spin liquid phase. For another iridate H$_3$LiIr$_2$O$_6$ which does not order to very low temperatures and shows anomalous gapless behavior quite distinct from the Kitaev model,\cite{Takagi} competing theories exists in terms of Majorana cones\cite{Slagle2018} and random singlet phases.\cite{Itamar2} Here, noise-correlations can be a very useful tool for mapping out the structure factor and therefore figuring out whether the ground state is a quantum spin liquid or not. 

Triangular lattice insulating organic compounds like $\kappa$-Et or Pd-dmit have also been proposed as quantum spin liquid candidates.\cite{Yamashita,Yamashita2,Yamashita3} In-plane thermal transport experiments in these materials show strong evidence of exotic gapless excitations which do not carry electric charge. Measuring noise correlations via magnetometry can provide strong evidence in favor of these elementary excitations carrying a fractionalized spin of half, and therefore a spin liquid ground state.

Another class of compounds include antiferromagnets on the highly frustrated Kagome lattice, like the intensely studied Herbertsmithite\cite{Mendels,Jeschke,Olariu,Helton,deVries,Bert,han2012} and Kapellasite.\cite{Wills2012} While Herbertsmithite appears to be gapped,\cite{Han2016} Kapellasite seems to be gapless with the precise nature of the ground state still unknown.\cite{Wills2012}  Therefore, our study of noise in gapless spin liquids of different kinds is very relevant for Kapellasite (and possibly other frustrated Kagome compounds).

There is an ongoing debate over the precise nature of the low-temperature paramagnetic states in compounds like YbMgGaO$_4$ and YbZnGaO$_4$. The $T^{0.7}$ specific heat in YbMgGaO$_4$ is reminiscent of the RPA calculation in a spinon Fermi surface as such a QSL is expected to $T^{2/3}$ specific heat.\cite{Shen2016,LiLuChen} 
Reference~\onlinecite{Itamar1}, instead, argues that the $0.7$ exponent is coincidental and the phase is actually described the random singlet model discussed above, and the same phase describes YbZnGaO$_4$ with an exponent $0.59$. As we saw, these two phases have very different signatures in the magnetic noise, and therefore noise magnetometry can serve as a probe that resolves the actual nature of the paramagnetic phase in these compounds.

In the frustrated $S = 1$ triangular lattice compound Ba$_3$NiSb$_2$O$_9$, several proposals exist for the ground state, including a putative spin liquid,\cite{CHR2012} quadratic band touching of spinons\cite{XWYBF_2012} and a spinon Fermi surface.\cite{Fak_QSL2017} These phases have distinct signatures in the magnetic noise and hence studing the relaxation time can be used to distinguish these candidate phases. 

Regarding observation of anyonic statistics, the most likely candidate would be a chiral spin liquid state. These phases have recently been observed in a DMRG study of the Hubbard model on the triangular lattice,\cite{szasz2018} raising hopes of finding such a ground state in the organic insulators discussed previously. The relaxation time provides a  non-invasive route to measure the braiding statictics in such a phase. For fractional quantum hall states,\cite{Stern_FQH_Review} spectroscopic methods like measurement of local electronic density of states,\cite{Papic2018} have been suggested to detect anyonic statistics. Since the elementary anyonic excitations carry electric charge, the long-range unscreened Coulomb interaction between anyons are expected to strongly modify the threshold spectral function, and the effect of long-range interactions on the relaxation time is an interesting open problem left to future work. 

Recent proposals suggest the use of quantum impurities to study spin diffusion and magnon condensation in insulators, and image antiferromagnetic domain walls.
\cite{FT2018,AFdomainWall_Flebus} The first study is related to ours, and it is restricted to the spin-diffusive regime in magnetically ordered states, where two-magnon processes dominate over single-particle ones. As we argued, for small external  fields one would expect single magnons to dominate the magnetic fluctuations in an ordered state down to the lowest energy scales, whereas for a spin liquid this would no longer hold true. Hence, a clear distinction between these two phases can be diagnosed via spin qubit magnetometry.

\section{Conclusions and outlook}
\label{sec:conc}

The possibility to sample spin correlations in a wide range of energy and length scales make spin qubits an invaluable tool to probe two-dimensional magnetic insulators. We found that the probe-frequency, sample-probe distance and temperature dependence of the spin relaxation time can furnish valuable information about the nature of the phase in gapless systems. Given the large number of experimental candidates for exotic spin liquid phases, this minimally invasive technique holds great promise as a diagnostics of ground states. 
Further, spin qubits 
can also detect anyonic statistics in gapped systems which have been difficult to identify via more traditional probes. 
Noise magnetometry with single spin qubits, therefore, can open up new vistas for probing exotic phases of matter. 

\section*{Acknowledgments}
SC is grateful to I. Kimchi for valuable discussions and a detailed explanation of Ref.~\onlinecite{Itamar1}. We also thank E. Berg, C. Du, S. C. Morampudi, A. A. Patel, S. Sachdev, S. Whitsitt, A. Yacoby, N. Yao, Y. You, M. Zaletel and C. Zu for helpful discussions. SC acknowledges support from the NSF under Grant No. DMR-1664842. JFRN and ED acknowledge support from Harvard-MIT CUA, NSF Grant No. DMR-1308435 and AFOSR-MURI: Photonic Quantum Matter (award FA95501610323). 

\begin{widetext}

\appendix
\section{relaxation time for magnetic insulators}
\label{app:RT}
\subsection{Relaxation rate of the spin probe}
In this appendix, we compute the relaxation time-scale of the spin probe in response to magnetic field fluctuations, using Fermi's Golden rule. Recall that the probe Hamiltonian is given by:
\beq
\mathcal{H} = \frac{\omega}{2} \sigma_z + \mu_B ~ \bm{\sigma} \cdot \bm{B}(\r,t),
\eeq
where $\bm{B}(\r,t)$ represents the time-varying magnetic field at the location of the sample. We assume that the back-reaction of the probe spin on the sample can be neglected, and that the sample is in thermal equilibrium at temperature $T = \beta^{-1}$. Denoting the eigenstate of the sample and spin polarization probe by the product state $\ket{n,\sigma} = \ket{n} \otimes \ket{\sigma}$ (with energy $\varepsilon_n$), we have the following emission rate of the probe initially prepared in the $\ket{+}$ state.

\begin{eqnarray}
R_{\rm em} &=& 2 \pi \sum_{n,m} \frac{e^{-\beta \on}}{Z} \big| \bra{m,-}  \mu_B  \bm{\sigma} \cdot \bm{B} \ket{n,+} \big|^2 ~ \delta(\omega + \varepsilon_n - \varepsilon_m) \nn
& = & 2 \pi ( \mu_B)^2 \sum_{n,m} \frac{e^{-\beta \on}}{Z} \left[ B^x_{nm} B^x_{mn} + B^{y}_{nm}B^{y}_{mn} + i B^y_{nm} B^x_{mn} - i B^x_{nm} B^y_{mn} \right] \delta(\omega + \varepsilon_{nm}) \nn
& = & 2 \pi ( \mu_B)^2 \sum_{n,m} \frac{e^{-\beta \on}}{Z} B^-_{nm} B^{+}_{mn} \delta(\omega + \varepsilon_{nm}), \text{ where } B^{\pm} = B^x \pm i B^y, 
\end{eqnarray}

where $B^j_{nm} = \bra{n} B^j \ket{m}$, and $\varepsilon_{nm} = \varepsilon_n - \varepsilon_m$. Note that only the mode of the magnetic field $\B$ oscillating at frequency $\omega$ couples to the probe, so the application of Fermi's Golden rule is justified. Similarly, the emission rate is given by the following expression:
\beq
R_{\rm abs} &=&  \pi ( \mu_B)^2 \sum_{n,m} \frac{e^{-\beta \varepsilon_n}}{Z} \left[ B^x_{nm} B^x_{mn} + B^{y}_{nm}B^{y}_{mn} - i B^y_{nm} B^x_{mn} + i B^x_{nm} B^y_{mn} \right] \delta(\omega - \varepsilon_{nm}) \nn
& = &   \pi ( \mu_B)^2 \sum_{n,m} \frac{e^{-\beta \varepsilon_n}}{Z} B^+_{nm} B^-_{mn} \delta(\omega - \varepsilon_{nm}). 
\eeq
The relaxation rate is defined as the average of the absorption and emission rates, $T^{-1}_{1} = \frac{1}{2}[R_{\rm abs} + R_{\rm em}]$, and it can be expressed conveniently in terms of the noise tensor $\N_{ij}(\omega)$ defined as follows:
\beq
\N_{ij}(\omega) = \frac{1}{2} \int_{-\infty}^{\infty} dt \, \langle \{ B^i(t), B^j(0) \} \rangle e^{i \omega t} = \sum_{n,m} \frac{e^{-\beta \varepsilon_n}}{Z} \left[  B^i_{nm} B^j_{mn} \delta(\omega + \varepsilon_{nm})  + B^j_{nm} B^i_{mn} \delta(\omega - \varepsilon_{nm}) \right] \nn
\label{eq:noiseTens}
\eeq
By comparing Eq.~(\ref{eq:noiseTens}) with $1/T_{1}$, we see that the following expression holds:
\beq
\frac{1}{T_{1}} = ( \mu_B)^2 \N_{-+}(\omega).
\label{eq:TinvApp}
\eeq
Using the fluctuation-dissipation theorem (which can be proven using spectral representations), we can re-write the noise tensor in terms of the spectral density of the magnetic field.
\beq
\N_{ij}(\omega) =\frac{1}{2} \coth\left( \frac{\omega}{2T} \right) \S_{ij}(\omega), \text{ where } \S_{ij}(\omega) =   \int_{-\infty}^{\infty} dt \, \langle [ B^i(t), B^j(0) ] \rangle e^{i \omega t}
\eeq
Further, we can also write the spectral density in terms of the retarded correlators of the magnetic field, which are more convenient to calculate. 
\beq
\S_{ij}(\omega) = -  \text{Im}[ C^R_{B^i B^j}(\omega)], \text{ where } C^R_{B^i B^j}(\omega) = -i \int_{-\infty}^{\infty} dt \, \Theta(t) \langle [ B^i(t), B^j(0) ] \rangle e^{i \omega t}
\eeq

\subsection{Sample-induced magnetic fluctuations}

In the main text, we used the dipole approximation (neglecting retardation effects) to calculate the magnetic field fluctuations at the probe location to the thermal spin fluctuations in the sample. In this appendix, we obtain the same by a more elementary approach, i.e, directly solving Maxwell's equations. Recall that Maxwell's equations in Lorentz gauge are given by ($\mu_B = e/2m_e$ is the Bohr Magneton, $\hbar = 1$):
\beq
\partial^2 A^{\mu} =  \left(- \frac{\partial_t^2}{c^2} + \nabla^2 \right) A^{\mu} = \mu_0(0, \nabla \times \m)^{\mu}, \text{ where } \m(\bm\rho,z,t) = - g_\sigma \mu_B \bS(\bm\rho,t) \delta(z) \nn
\eeq
where we have set the lattice spacing $a = 1$. Let us first define the magnetic kernel $G^{\mu}_i$ as follows (with Einstein summation on repeated indices implied):
\beq
A^\mu(\r, t) = 
\mu_0 \int dt^\prime \, d\rp \, G^{\mu}_i (\r - \rp, t - t^\prime) m_i(\rp, t^\prime)
\label{eq:Amu}
\eeq
where $G^{\mu}_i (\r - \rp, t - t^\prime)$ satisfies the following differential equation:
\beq
 \left(- \frac{\partial_t^2}{c^2} + \nabla^2 \right) G^{\mu}_i (\r - \rp, t - t^\prime) =  \delta(t - t^\prime)\left[ 0, \nabla \times (\delta(\bro - \bro^\prime) \delta(z - z^\prime) \hat{e}_i) \right]^\mu
\label{eq:Gf}
\eeq
We now specialize to a translation invariant phase of the sample of area $L^2$, governed by a time-independent Hamiltonian (note that this is justified because we do not consider back-reaction of the probe). Then, we can re-write Eqs.~(\ref{eq:Amu}) and the Green's function in and (\ref{eq:Gf}) in terms of Fourier modes as follows:
\beq
A^\mu(\r, t) = \frac{1}{\sqrt{L^2}} \sum_{\q} \int \frac{d\omega}{2\pi} \; A^{\mu}(z, \q, \omega) e^{i (\q \cdot \bro - \omega t)}, &~&G^\mu_i(\r, t) = \frac{1}{L^2} \sum_{\q} \int \frac{d\omega}{2\pi} \; G^{\mu}_i(z, \q, \omega) e^{i (\q \cdot \bro - \omega t)} \nn
m_i(\r, t) = \frac{1}{\sqrt{L^2}} \sum_{\q} \int \frac{d\omega}{2\pi} \; m_i(\q, \omega) e^{i (\q \cdot \bro - \omega t)} \delta(z)
\eeq
Plugging these into Eq.~(\ref{eq:Gf}), we end up with the following equation:
\beq
\left(-\lambda^2 + \partial_z^2 \right) G^{\mu}_i(z,\q,\omega) = [0, (i q_x, i q_y, \partial_z) \times (\delta(z) \hat{e}_i)]^\mu && \text{ where } \lambda = \sqrt{\q^2 -  \frac{\omega^2}{c^2}}
\label{eq:GfF}
\eeq
The solutions to Eq.~(\ref{eq:GfF}) can be written as follows:
\beq
G^{\mu}_x(z,\q,\omega) = \frac{e^{- \lambda |z|}}{2} \begin{pmatrix}
0 \\
0 \\
\sgn(z)  \\
\frac{i q_y}{\lambda} 
\end{pmatrix}^\mu, ~ G^{\mu}_y(z,\q,\omega) = \frac{e^{- \lambda |z|}}{2} \begin{pmatrix}
0 \\
-\sgn(z)  \\
0 \\
-\frac{i q_x}{\lambda} 
\end{pmatrix}^\mu, ~ G^{\mu}_z(z,\q,\omega) = \frac{e^{- \lambda |z|}}{2} \begin{pmatrix}
0 \\
\frac{- iq_y}{\lambda} \\
\frac{iq_x}{\lambda} \\
0 \\
\end{pmatrix}^\mu \nn
\eeq
Note that these are consistent with our choice of Lorentz gauge, for which a sufficient condition is $\partial_{\mu} G^{\mu}_i = 0$ for each $i$. We use the vector potential to find the magnetic field $\B(\r,t)$ at the location of the probe by $\B(\r,t) = \nabla \times \bm{A(\r,t)}$, 
\beq
\B(\r, t) = \frac{\mu_0}{\sqrt{L^2}} \sum_{\q} \int \frac{d\omega}{2\pi} ~ \bm{H}_{i}(z,\q,\omega)  e^{i (\q \cdot \bro - \omega t)} m_i(\q,\omega)
\label{eq:B}
\eeq
where $\bm{H}_{i}(z,\q,\omega)  =  (i q_x, i q_y, \partial_z) \times \bm{G}_i(z, \q, \omega)$ is given by (here we choose $z>0$):
\beq
\bm{H}_x = \frac{e^{-\lambda z}}{2} \left( \frac{\lambda^2 - q_y^2}{\lambda}, \frac{q_x q_y}{\lambda}, iq_x \right), ~ \bm{H}_y = \frac{e^{-\lambda z}}{2} \left(  \frac{q_x q_y}{\lambda}, \frac{\lambda^2 - q_x^2}{\lambda},iq_y \right), ~  \bm{H}_z = \frac{e^{-\lambda z}}{2} \left( i q_x, i q_y, - \frac{\q^2}{\lambda} \right) \nn
\eeq
Finally, we plug the resultant expression into Eq.~(\ref{eq:TinvApp}) to get the relaxation rate for a probe initially polarized in the $\ket{+}$ direction. The magnetic field correlators can in turn be expressed in terms of the kernels $\bm{H}_i$ and the magnetization correlators in the sample, using the form of $\B(\r, t)$ from Eq.~(\ref{eq:B}). We can take advantage of translation invariance of the sample in the x-y plane and time independence of the sample Hamiltonian to make the following simplifcation for the magnetization correlators in the sample. 
\beq
\langle [m_i(\q_1,\omega_1), m_j(\q_1,\omega_2)] \rangle =2\pi \delta(\omega_1 + \omega_2) \delta_{\q_1, - \q_2} \langle [m_i(\q,\omega), m_j(-\q,-\omega)] \rangle  
\eeq
After this simplication, we find that we can express the correlator as:
\beq
\S_{-+}(\omega) = \frac{1}{2L^2} \sum_{\q} ~ \bm{H}^-_{i}(z,\q,\omega) \bm{H}^+_{j}(z,-\q, -\omega) \langle [m_i(\q,\omega), m_j(-\q,-\omega)] \rangle \nn
\eeq
Note that we can write the correlator as follows (schematically, with the $\q, \omega$ dependences implicit):
\beq
 \bm{H}^-_{i}\bm{H}^+_{j} \langle [m_i, m_j] \rangle &=& \bigg\langle \left[\frac{1}{2}(H^-_+m_- + H^-_- m_+) + H^-_z m_z, \frac{1}{2}(H^+_+m_- + H^+_- m_+) + H^+_z m_z \right] \bigg \rangle \nn
 & = & \frac{1}{4} \left( H^-_+H^+_- \langle [m_- , m_+] \rangle + H^-_-H^+_+ \langle [m_+ , m_-] \rangle \right) + H^+_z H^-_z \langle [m_z , m_z] \rangle + ... \nn
\label{eq:Simp}
\eeq
The terms included in the ellipsis have zero matrix element if the total $S_z$ commutes with the sample Hamiltonian, so that the many-body eigenstates $\ket{n}$ also have fixed $S_z$. Alternatively, because of their form, these terms integrate to zero during the momentum integration provided the spin-correlators ${\cal C}_{\alpha\beta}(\q,\omega) =  \langle [S^\alpha(\q, \omega) , S^\beta(-\q,-\omega)] \rangle$ depend only on $q$, i.e, the low-energy theory has rotational symmetry about $\q = 0$. Even if they do not vanish, they will not make any qualitative difference to the relaxation-rate, so we will neglect these terms. 

Finally, we calculate the products of the Kernels shown schematically in Eq.~(\ref{eq:Simp}), and make another simplifying approximation: $\omega/c \ll q$ in most condensed matter systems (equivalent to taking the speed of light to be infinite) so that $\lambda \approx q$:
\beq
H^+_+(\q, \omega) H^-_-(-\q, -\omega) &=& \left( \frac{e^{-\lambda z}}{2 \lambda} \right)^2 q^4 \approx \frac{q^2 e^{- 2 q z}}{4} \nn
H^+_-(\q, \omega)  H^-_+(-\q, -\omega)  &=&  \left( \frac{e^{-\lambda z}}{2 \lambda} \right)^2 (2 \lambda^2 - \q^2)^2 \approx \frac{q^2 e^{- 2 q z}}{4} \nn
H^+_z(\q, \omega)  H^-_z(-\q, -\omega) & = &  \left( \frac{e^{-\lambda z}}{2} \right)^2 q^2 \approx \frac{q^2 e^{- 2 q z}}{4}
\eeq
Plugging these back into the expression for the spectral function $\S_{ij}(\omega)$ for the magnetic field and setting $z = d$ and $g_\sigma = 2$, we arrive at Eq.~(\ref{eq:T12}), which is reproduced below for convenience.
\beq
\frac{1}{T_{1}} &=& 4(\mu_0 \mu_B)^2 \coth\left( \frac{\omega}{2T} \right) \frac{1}{L^2} \sum_{\q} F(d, \q) \, \text{Im} \left[ -\frac{1}{4} \left( \mathcal{C}_{-+}(\q, \omega) + \mathcal{C}_{+-}(\q, \omega) \right) - \mathcal{C}_{zz}(\q, \omega)  \right] \nn
& \xrightarrow{L \rightarrow \infty} & (\mu_0 \mu_B)^2 \coth\left( \frac{\omega}{2T} \right) \int \frac{d^2q}{(2\pi)^2} F(d, \q) \, \text{Im} \left[ - \left( \mathcal{C}_{-+}(\q, \omega) + \mathcal{C}_{+-}(\q, \omega) \right) - 4 \mathcal{C}_{zz}(\q, \omega)  \right] 
\eeq
where $F(d, \q) = q^2 e^{- 2 q d}/8 \approx  \sum_{i = x,y,z} |\bm{H}_{i}(d,\q,\omega)|^2 / 2$ is a distance dependent form factor that shows that the integral over the Brillouin Zone is dominated by $q \sim d^{-1}$, and $\mathcal{C}_{ij}(\q, \omega)$ are the retarded spin-spin correlations that depend solely on the equilibrium fluctuations of magnetization of the sample. Note that here we have also made the approximation that the typical velocity scale of propagation of excitations in the sample is much smaller than $c$, and therefore $\lambda \approx q$ and $F(d, \q)$ is independent of $\omega$. 

\section{Computations of relaxation times}
\subsection{Gapless systems}
\label{app:integrals2}
For the clean Dirac spin liquid in the extended Kitaev honeycomb model in Eq.~(\ref{eq:kitaevhgamma}), the spin operator can be written down explicitly in terms of the low-energy Dirac fermions\cite{SYB2016} as
\beq
\sigma^a = \psi^\dagger m^a \psi  + ( i \psi^T \n^{\mu,a} \cdot \nabla \psi \, e^{- i \bm{K} \cdot \r} + \mbox{H.c.}),
\eeq
where $m^a$ and $\n^{\mu,a}$ are two-by-two diagonal matrices. The primary contribution to the low-energy structure factor comes from the first term which is more relevant (the second term has a derivative). For simplicity, here we just consider $m^a = \sigma^0$ and use existing results for density-density correlations for Dirac fermions in graphene where the dispersion is identical (although graphene has two flavors of fermions as opposed to a single one here). Further, there is no long-range Coulomb interaction for fermionic spinons, and the gapped $\mathbb{Z}_2$ gauge field mediates a weak short range interaction which is irrelevant. So we can use the bare susceptibility from Ref.~\onlinecite{AVAPM2009} by setting the chemical potential $\mu = 0$ as we are interested in the case with zero doping.
\beq
\mathcal{C}^{~''}_{+-}(\q,\omega) &=& \Theta(vq - \omega) \frac{q^2}{\sqrt{v^2q^2 - \omega^2}} [G_-(\q,\omega,T) - G_+(\q, \omega,T)] \nn 
&& + \Theta(\omega - vq)  \frac{q^2}{\sqrt{ \omega^2 - v^2q^2}} \left[\frac{\pi}{2} - 2 H_+(\q,\omega,T) \right], \text{ where } \nn
G_{\pm}(\q, \omega, T) &=& \int_{1}^{\infty} du \frac{\sqrt{u^2 - 1}}{e^{(|vqu \pm \omega|)/2T}+ 1}, ~ H_{\pm}(\q, \omega, T) = \int_{-1}^{1} du \frac{\sqrt{1-u^2 }}{e^{(|vqu \pm \omega|)/2T}+ 1}
\eeq
We note that for $T \rightarrow 0$, this is reduced to the form we have in Eq.~(\ref{eq:SCleanDirac}) as both $G_{\pm}$ and $H_+$ go to zero in this limit. Therefore the relaxation time in the limit $T \ll\omega$ is given by:
\beq
\frac{1}{T_1} & \propto &  \int_0^{\omega/v} dq \, q^3 e^{- 2 q d} \, \frac{q^2}{\sqrt{\omega^2 - v^2 q^2}} \nn
& \approx &  \begin{cases}
\omega^5 , ~ d \omega/v \ll 1 \\
\frac{1}{\omega d^6}, ~  d \omega/v  \gg 1
\end{cases}
\label{eq:DT0TempCDirac}
\eeq
For very large $T \gg \omega >0$, we can again find a somewhat simple expression for the structure factor by approximating the Fermi-Dirac distribution by the Boltzmann distribution. 
\beq
\mathcal{C}^{~''}_{+-}(\q,\omega) & \approx & \Theta(vq - \omega) \, \frac{q^2}{\sqrt{v^2 q^2 - \omega^2}} \left[ 2 \sinh\left(\frac{\omega}{2T}\right) \frac{2T}{vq} K_1\left( \frac{vq}{2T} \right) \right] \nn
&& + \Theta(\omega - vq) \, \frac{q^2}{\sqrt{\omega^2 - v^2 q^2}} \frac{\pi}{2} \left[  1- e^{-\omega/2T}  \right]
\label{eq:SFcleanDirac}
\eeq
In this limit, we may again calculate the relaxation rate to leading order in $\omega/T$. Note that we can also define an effective temperature scale $\T_d = \hbar v/k_B d$, restoring the fundamental constants for clarity. As discussed in the main text, both $\T_d \gg T$ and $\T_d \ll T$ ar experimentally accessible limits. However, for our calculations we stick to the regime $\T_d \ll T$, i.e, the temperature is the largest energy scale in the problem. In this regime, we can approximate $\coth(\omega/2T) \approx 2T/\omega$, and extract analytical expressions for the relaxation rates in the regimes $\omega \ll \T_d$ and $\omega \gg \T_d $.
\beq
\frac{1}{T_1} & \propto & \frac{2T}{\omega}  \int_0^{\infty} dq \, q^3 e^{- 2 q d} \, \mathcal{C}^{~''}_{+-}(\q,\omega,T)  \nn
& \approx &  \begin{cases}
\frac{T^2}{d^3}, ~  d \omega/v  \ll 1 \\
\frac{T^2 \omega^{5/2}}{\sqrt{d}} e^{-2\omega d/v}, ~  d \omega/v  \gg 1
\end{cases}
\label{eq:DT0FreqCDirac}
\eeq

For the Dirac spin liquid at finite doping, we again use the susceptibility from Ref.~\onlinecite{AVAPM2009} at finite chemical potential $\mu$. We also assume that $\mu$ is the largest energy scale in the problem, so that the temperature $T$, the probing frequency $\omega$ and the temperature scale $\T_d$ set by the inverse distance $d$ are all much less that $\mu$. 
\beq
\mathcal{C}^{~''}_{+-}(\q,\omega) &\approx& \sum_{\alpha = \pm} \Theta(v q - \omega) \frac{q^2}{\sqrt{v^2q^2 - \omega^2}} \left[ G^\alpha_-(\q, \omega, T) - G^\alpha_+(\q, \omega, T) \right] \nn
&& ~~~~~~~ + \Theta(\omega - vq) \frac{q^2}{\sqrt{\omega^2 - v^2q^2}}\left[ \frac{\pi}{2}\delta_{\alpha,-} - H^\alpha_+(\q, \omega, T)\right], \text{ where } \nn
G^\alpha_{\pm}(\q, \omega, T) &=& \int_{1}^{\infty} du \frac{\sqrt{u^2 - 1}}{e^{(|vqu \pm \omega| - 2 \alpha \mu)/2T}+ 1}, ~ H^\alpha_{\pm}(\q, \omega, T) = \int_{-1}^{1} du \frac{\sqrt{1-u^2 }}{e^{(|vqu \pm \omega| - 2 \alpha \mu)/2T}+ 1}. \nn
\label{eq:SFdopedDirac}
\eeq
Note that for $\mu \gg T$, the integrals in $G^\alpha$ and $H^\alpha$ contribute to the correlation function appreciably only for $\alpha = +$. First, we look at the limit of $\omega \gg T$, whence we can replace the Fermi functions by theta functions for the integrals in $G^-_{\pm}$, and we have:
\beq
G^+_-(\q, \omega, T) - G^+_+(\q, \omega, T)   &\xrightarrow{T \rightarrow 0}&  \int_{(2\mu - \omega)/vq}^{(2\mu + \omega)/vq} du \, \sqrt{u^2 - 1} \approx \frac{\omega}{vq} \left( \frac{\mu}{vq}\right)^2, \text{ and }  \nn
H^+_+(\q, \omega, T) &\xrightarrow{T \rightarrow 0}& \int_{-1}^{1} du \sqrt{1-u^2 } = \frac{\pi}{2}, \nn
\eeq
where we have used that  the theta function imposed upper limit $\frac{2\mu + \omega}{vq} \gg 1$. Therefore, up to corrections exponentially suppressed by $e^{-\mu/T}$, the term proportional to $\Theta(\omega - vq)$ in Eq.~(\ref{eq:SFdopedDirac}) does not contribute. So the relaxation time in the limit $T \ll \omega \ll \mu$ is given by:
\beq
\frac{1}{T_1} & \propto &  \int_{\omega/v}^\infty dq \, q^3 e^{- 2 q d} \, \frac{q^2}{\sqrt{v^2 q^2 - \omega^2}} \frac{\omega \mu^2}{(vq)^3} \approx \frac{\mu^2\omega^2}{2d}\left[ \frac{2d\omega}{v} K_0(2d \omega/v) + K_1(2d\omega/v)\right]  \nn
& \approx &  \begin{cases}
\frac{\omega}{d^2} , ~ d \omega/v \ll 1 \\
\frac{\omega^{5/2}}{\sqrt{d}} e^{-2\omega d/v}, ~  d \omega/v  \gg 1.
\end{cases}
\eeq
Now we study the other limit where $\omega \ll T \ll \mu$, where we have the following limiting form of $G^\alpha_\beta$:
\beq
G^+_-(\q, \omega, T) - G^+_+(\q, \omega, T) &=& \int_{1}^{\infty} du \, \sqrt{u^2-1} \, \frac{e^{(vqu - 2\mu)/2T} \sinh(\omega/2T)}{e^{(vqu - 2\mu)/T} + e^{(vqu - 2\mu)/2T} \cosh(\omega/2T) + 1} \nn
& \xrightarrow{T \gg \omega} & \sinh\left(\frac{\omega}{2T}\right) \int_{1}^{\infty} du \, \frac{\sqrt{u^2-1}}{4\cosh^2[(vqu - 2\mu)/2T]} \nn &\approx & \sinh\left(\frac{\omega}{2T}\right) \left(\frac{\mu}{vq}\right)^2
\eeq
Note that $H^+_+ \approx \pi/2$ still holds upto corrections of O$(e^{-\mu/T})$, and hence  the term proportional to $\Theta(\omega - vq)$ in Eq.~(\ref{eq:SFdopedDirac}) again does not contribute. Therefore, the relaxation time in the limit $ \omega \ll T \ll \mu$ is given by:
\beq
\frac{1}{T_1} & \propto &  \int_{\omega/v}^\infty dq \, q^3 e^{- 2 q d} \, \frac{q^2}{\sqrt{v^2 q^2 - \omega^2}} \frac{\mu^2}{(vq)^2} \sim \frac{\mu^2\omega^2}{2d}\left[ \frac{2d\omega}{v} K_1(2z \omega/v) + K_2(2d\omega/v)\right]  \nn
& \approx &  \begin{cases}
\frac{1}{d^3} , ~ d \omega/v \ll 1 \\
\frac{\omega^{5/2}}{\sqrt{d}} e^{-2\omega d/v}, ~  d \omega/v  \gg 1
\end{cases}
\eeq

In case of time-reversal symmetry preserving disorder for the Kitaev spin liquid, the disorder term appears like a vector potential in the low-energy Hamiltonian, which we assume to have short range correlations. 
\beq
H = \sum_{\k, \k^\prime} \psi^\dagger_{\k} \left( v \, \bm{\sigma} \cdot \k \, \delta_{\k,\k^\prime} + \bm{A}_{\k - \k^\prime} \cdot \bm\sigma \right) \psi_{\k^\prime}, ~~ \langle \bm{A}_{\q} \bm{A}_{\q^\prime} \rangle = (2\pi)^2 \delta(\q + \q^\prime) \Delta_A
\eeq
The properties of this system has been studied in detail in Ref.~\onlinecite{Ludwig}, so we use their results to determine the scaling of the structure factor, and hence, the relaxation time. Let us review a few key results from Ref.~\onlinecite{Ludwig}, which we will use extensively. At $\omega = 0$, the system is described by a fixed line of interacting 1+1 d theories, characterized by $\omega/T$ scaling and a dynamical critical exponent z given by $\tz = 1 + \Delta_A/\pi$, where $\Delta_A$ is the disorder strength. The frequency $\omega$ (or energy) corresponds to a relevant operator with scaling dimension z, i.e, under scaling $q \rightarrow q/b$ and $\omega \rightarrow \omega /b^\tz$.

To calculate the structure factor, we can expand the fermionic spinon operators in single particle eigenstates for a fixed realization of disorder (neglecting interactions). Here we neglect the sublattice index for notational simplicity, one can put it back and check that it does not make any qualitative difference to the correlations at small momenta. 
\beq
\psi(\bro) = \sum_{\lambda} \phi_{\lambda}(\bro) f_{\lambda}
\eeq
Using these eigenstates, we shall evaluate the disorder averaged density-density correlator for the Dirac fermions to find the dynamic spin structure factor. Here we are assuming that the physical spin operator is $\sigma_a \sim \psi^\dagger m_a \psi$ with $m_a \sim \sigma_0$, as discussed for the free case. The density operator can be written as:
\beq
\rho(\q, i \omega_n) = \int d\bro \, e^{i \q \cdot \bro} \psi^\dagger(\bro, i \omega_n) \psi(\bro, i \omega_n) = \int d\bro \, e^{i \q \cdot \bro} \sum_{\lambda, \lambda^\prime} \phi_\lambda^*(\bro) \phi_{\lambda^\prime}(\bro) f^\dagger_{\lambda} f_{\lambda^\prime}
\eeq
Therefore, the susceptibility in the density channel in imaginary time is given by the thermal and disorder average of the density-density correlator, under the assumption that the system is self-averaging.
\beq
\chi(\q, i \omega_n) &=& - \frac{1}{\beta L^2} \Big\langle \rho(\q, i\on) \rho(-\q, -i\on) \Big\rangle_{thermal, disorder} \nn
& = & - \frac{1}{L^2} \int d\bro \int d\bro^\prime \, e^{i \q \cdot (\bro - \bro^\prime)} \sum_{\lambda, \eta} \frac{n_F(\xi_{\lambda}) - n_F(\xi_\eta)}{i \on + \xi_\eta - \xi_\lambda} \Big\langle \phi_\lambda^*(\bro) \phi_{\eta}(\bro) \phi_\eta^*(\bro^\prime) \phi_{\lambda}(\bro^\prime) \Big\rangle_{disorder} \nn
& = &  - \int d\bro \, e^{i \q \cdot \bro} \sum_{\lambda, \eta} \frac{n_F(\xi_{\lambda}) - n_F(\xi_\eta)}{i \on + \xi_\eta - \xi_\lambda} \Big\langle \phi_\lambda^*(\bro) \phi_{\eta}(\bro) \phi_\eta^*(0) \phi_{\lambda}(0) \Big\rangle_{disorder}
\label{eq:chiDisA1}
\eeq
where in the last step we assumed that disorder averaging restores translation invariance. Let us introduce the following function to simplify the calculation (further assuming rotational symmetry after disorder averaging):
\beq
g(q, \varepsilon, \varepsilon^\prime) = \sum_{\lambda, \eta} \int d\bro \, e^{i \q \cdot \bro}\Big\langle \phi_\lambda^*(\bro) \phi_{\eta}(\bro) \phi_\eta^*(0) \phi_{\lambda}(0) \Big\rangle_{disorder} \delta(\varepsilon - \xi_{\lambda}) \delta(\varepsilon - \xi_\eta)
\eeq
In terms of $g(q, \varepsilon, \varepsilon^\prime)$, we can rewrite the retarded correlator after analytically continuing Eq.~(\ref{eq:chiDisA1}) to real frequencies.
\beq
\chi(\q, \omega) &=& \int d\varepsilon \; d\varepsilon^\prime  \frac{n_F(\varepsilon) - n_F(\varepsilon^\prime)}{\omega + i0^+ + \varepsilon - \varepsilon^\prime} \; g(q, \varepsilon, \varepsilon^\prime) \nn 
\implies -\frac{1}{\pi}\text{Im}\left[ \chi(\q, \omega) \right]  &=& \int d\varepsilon \; [n_F(\varepsilon) - n_F(\varepsilon + \omega)] g(q, \varepsilon, \varepsilon + \omega) \approx \omega \int d\varepsilon \; \left( -\frac{dn_F}{d\varepsilon} \right) g(q, \varepsilon, \varepsilon) \nn
\label{eq:chiDisA2}
\eeq
In the last step, we made a low-energy approximation assuming $\omega$ to be the smallest energy scale, i.e $\omega \ll T$. Now, we need to find the scaling behavior of $g(q, \varepsilon_1, \varepsilon_2)$.
\beq
g(q, \varepsilon_1, \varepsilon_2) = b^{-y} g(bq, b^\tz\varepsilon_1, b^\tz\varepsilon_2)
\label{eq:gscaling}
\eeq
This can be done by comparing the expression for $\chi(\q, i\on)$ in the limit $\on \rightarrow 0$ from Eq.~(\ref{eq:chiDisA2}) with an alternate derivation of the static ($\omega_n = 0$) limit of the susceptibility from the knowledge of the scaling dimension of $\rho(\bro, \tau)$ in Ref.~\onlinecite{Ludwig}. 
\beq
\chi(\q, i\omega_n = 0, T) & = & \int_0^\beta d\tau \int d\bro \, e^{i \q \cdot \bro} \Big\langle \rho(\bro, \tau) \rho(0, 0) \Big\rangle \nn
& = & T \sum_{m,m^\prime}  \int d\bro \, e^{i \q \cdot \bro} \, b^{-2(2-z)}  \Big\langle \rho_m(\bro/b) \rho_{m^\prime}(0) \Big\rangle \nn
& = & T b^{-2(2-z)} b^2 \int \frac{dx \, dy}{b^2} e^{i (b\q) \cdot (\bro/b)}  \Big\langle \rho_m(\bro/b) \rho_{m^\prime}(0) \Big\rangle \nn
& = & T b^{2(z-1)} \chi(b q, i\on = 0, b^\tz T) \nn
&= & T^{(2-\tz)/\tz} \Phi_1\left( \frac{q}{T^{1/z}} \right)
\eeq
where in the last step we chose $b = T^{-1/\tz}$ and $\Phi$ is some universal scaling function. From Eq.~(\ref{eq:chiDisA2}) we can see that the following scaling holds:
\beq
\chi(\q, \omega = 0) &=& T \int \frac{d\varepsilon}{T} \; \frac{d\varepsilon^\prime}{T}  \frac{n_F(\varepsilon) - n_F(\varepsilon^\prime)}{\frac{\omega}{T} + i0^+ + \frac{\varepsilon}{T} - \frac{\varepsilon^\prime}{T}} \; g(q, \varepsilon, \varepsilon^\prime) \nn  
& = & T^{1 + y/\tz} \int \frac{d\varepsilon}{T} \; \frac{d\varepsilon^\prime}{T}  \frac{n_F(\varepsilon) - n_F(\varepsilon^\prime)}{\frac{\omega}{T} + i0^+ + \frac{\varepsilon}{T} - \frac{\varepsilon^\prime}{T}} \; g(q/T^{1/\tz}, \varepsilon/T, \varepsilon^\prime/T) \nn
& = & T^{1 + y/\tz} \, \Phi_1\left( \frac{q}{T^{1/\tz}} \right) 
\label{eq:chiDisA3}
\eeq
where we have again used that the integral in the second to last step is dimensionless to cast the result in terms of the scaling function $\Phi_1$. Comparing Eq.~(\ref{eq:chiDisA2}) and Eq.~(\ref{eq:chiDisA3}), we find that $y = 2 - 2 \tz$. Now, we can evaluate the relaxation time in the limit $\omega \ll T$, using Eq.~(\ref{eq:chiDisA2}) again to extract the linear term in $\omega/T$ which cancels the divergence from $\coth(\omega/2T)$ in the limit $\omega \ll T$. 
\beq
\frac{1}{T_1} & \propto & \coth\left( \frac{\omega}{2T} \right) \frac{\omega}{T} \cdot T^{(2-\tz)/ \tz}  \int_0^\infty dq \, q^3 \, e^{-2 q d} \;  \Phi_1\left( \frac{q}{T^{1/z}} \right) \nn
& \approx & T^{(6-\tz)/ \tz} \, \Psi_1(d T^{1/\tz})
\eeq
where $\Psi_1(d T^{1/\tz})$ is another universal scaling function. The anomalous scaling of frequency becomes apparent in the scaling of the relaxation time with distance from the sample! 

We now check that we get back the previously obtained results for clean Dirac fermions in the limit of zero disorder. In this case, we have relativistic scaling of space and time, i.e, $\tz = 1$. From Eq.~(\ref{eq:SFcleanDirac}), we check that the universal function in the $\omega \rightarrow 0$ limit is given by $\Phi_1(q/T) = 2 K_1(q/2T)$. Accordingly, the integral over $q$ gives a function $\Psi_1(dT)\approx (dT)^{-3}$, which combined with the $T^5$ factor upfront for $\tz = 1$ reproduces the scaling of the relaxation time as $T^2/d^3$ in Eq.~(\ref{eq:DT0FreqCDirac}). The second limit in Eq.~(\ref{eq:DT0FreqCDirac}) when $d\omega/v \gg 1$ is difficult to capture by the scaling argument which naturally assumes $\omega$ to be the smallest energy scale, whereas $\T_d \ll \omega$ in this case. 

We can also study the $T \rightarrow 0$ limit by choosing $b = \omega^{-1/\tz}$ in Eq.~(\ref{eq:gscaling}). In this limit we find that:
\beq
\frac{1}{T_1} & \propto & \coth\left( \frac{\omega}{2T} \right) \cdot \omega^{(2-\tz)/ \tz}  \int_0^\infty dq \, q^3 \, e^{-2 q d} \;  \Phi_2\left( \frac{q}{\omega^{1/z}} \right) \nn
& \approx & \omega^{(6-\tz)/ \tz} \, \Psi_2(d \, \omega^{1/\tz})
\eeq
For the clean system, we can again put $\tz = 1$ and hence see that for $d \omega \ll 1$ we have $\Psi_2(d \, \omega) \sim 1$ for $\omega d/v \ll 1$ and $\Psi_2(d \, \omega) \sim (d \, \omega)^{-6}$ for $\omega d/v \ll 1$, reproducing the relaxation rates in Eq.~(\ref{eq:DT0TempCDirac}).

For the $\mathbb{Z}_2$ spin liquid with a spinon Fermi surface, we can study the relaxation time in the limit of $T \ll \omega \ll \mu$. In this regime, the density-density correlation of the fermion field is given by a diffusive form,\cite{Castellani1986} and therefore the spin structure factor also assumes a diffusive form:
\beq
\mathcal{C}^{~''}_{+-}(\q,\omega) \sim - \text{Im}\left[ \frac{\nu D_s q^2}{-i \omega + D_s q^2} \right] \approx \frac{\nu D_s q^2 \omega}{\omega^2 + D_s^2 q^4}
\eeq
In the of limit of small $T/\omega$, $ \coth(\omega/2T) \approx 1$, so the relaxation time is given by:
\beq
\frac{1}{T_1} \approx \int_0^\infty \, dq \, q^3 \, e^{-2qd} \frac{\nu D_s q^2 \omega}{\omega^2 + D_s^2 q^4}.
\eeq
If $d$ is small so that $\omega d^2 \ll D_s$, then the integral is essentially cutoff by the exponential factor at a scale of $q \sim d^{-1}$. Setting $D_s = 1$, we have
\beq
\frac{1}{T_1} \sim \int_0^\infty \, dq \, q^3 \, e^{-2qd} \frac{\nu  q^2 \omega}{\omega^2 +  q^4} \approx \int_0^{1/2d} \, dq \, q^3  \frac{\nu q^2 \omega}{\omega^2 + q^4} = \frac{1}{8} \left( \frac{\omega}{d^2} - 4 \omega^2 \cot^{-1}(4 d^2 \omega) \right) \approx \frac{\omega}{8 d^2}  \nn
\eeq
On the other hand, for large $d$ with $\omega d^2 \gg D_s$, the integrand is dominated by small $q \sim d^{-1}$ in the numerator, and the denominator can be assumed to be roughly $\omega^2$ for the regime where the exponential factor is small. 
\beq
\frac{1}{T_1} \sim \int_0^\infty \, dq \, q^3 \, e^{-2qd} \frac{\nu  q^2 \omega}{\omega^2 +  q^4} \approx \frac{1}{\omega} \int_0^\infty \, dq \, q^3 \, e^{-2qd} (\nu  q^2 ) \sim \frac{1}{\omega d^6}
\eeq

Finally, we arrive at the consideration of a clean U(1) spin liquid with a Fermi surface. In this case, the largest contribution comes from the imaginary part of the self-energy (that scales as $\omega^{2/3}$) in the RPA Green's function of the spinon.\cite{LeeNagaosa,Polchinski1994} In order to evaluate the susceptibility, we write the spinon Green's function in terms of  the spectral representation. 
\beq
G_f(\k, i \on) &=& \int_{-\infty}^{\infty} \frac{d\varepsilon}{2\pi} \frac{A_f(\k, \varepsilon)}{i \on - \varepsilon} , \nn \text{ where } 
 A_f(\k, \varepsilon) &=& - 2 \text{Im}[G_f(\k, \varepsilon)]  =
\begin{cases}
2\pi \delta(\xi_{\k}) \text{ if } \varepsilon = 0 \nn
\frac{2 \, C \, \varepsilon^{2/3}}{\xi_{\k}^2 + C^2 \varepsilon^{4/3}}, \text{ if } \varepsilon > 0
\end{cases},
\eeq
where $C \approx \mu^{1/3}$ is a constant. Writing in terms of the spectral function allows us to do the Matsubara summation in the density-density correlator. After some algebra, we can write the imaginary part of the retarded density-density correlator (after analytic continuation to real frequencies) as:
\beq
\mathcal{C}^{~''}_{+-}(\q,\omega) &=&  -\int_{-\infty}^{\infty} \frac{d\varepsilon}{2\pi} \int \frac{d\k}{(2\pi)^2} A(\k, \varepsilon) A(\k + \q, \omega + \varepsilon) \left[ n_F(\varepsilon + \omega) - n_F(\varepsilon) \right] \nn
& \approx &  \int_{-\infty}^{\infty} \frac{d\varepsilon}{2\pi} \int \frac{d\k}{(2\pi)^2} A(\k, \varepsilon) A(\k + \q, \omega + \varepsilon) \; \omega \left(-\frac{\partial n_F}{\partial \varepsilon}\right) \nn
& \approx & \omega \int \frac{d\k}{(2\pi)^2} A(\k, 0) A(\k + \q, \omega)
\eeq
where we have first assumed $\omega$ is small compared to $\mu$ to replace the difference in Fermi functions by a derivative, and then further used $T \ll \mu$ to approximate the Fermi function by a step function so that its derivative is a delta function. We can further simplify the integral in the low $q$ limit, which is reasonable to consider as typically $q \ll k_F$. We have $\xi_{\k  + \q} \approx v_F q \cos(\theta) + O(q^2)$ in this limit, where $\theta$ is the angle between $\k $ and $\q$.
\beq
 \int \frac{d\k}{(2\pi)^2} A(\k, 0) A(\k + \q, \omega) &=&  \int \frac{d\k}{2\pi} \, \delta\left( \frac{k^2}{2m} - \mu \right) \frac{2 \, C \, \omega^{2/3}}{\xi_{\k  + \q}^2 + C^2 \omega^{4/3}} \nn
& = & \frac{m \,}{\pi} \int d\theta \; \frac{ C \, \omega^{2/3}}{v_F^2 q^2 \cos^2\theta + C^2 \omega^{4/3}} \nn
& \propto & \frac{1}{\sqrt{v_F^2 q^2 + C^2 \omega^{4/3}}}
\eeq
Now, we can estimate the relaxation time in the $T \rightarrow 0$ limit. For $\omega \gg T$, we have $ \coth(\omega/2T) \approx 1$. The scaling of the relaxation time is then given by:
\beq
\frac{1}{T_1}& \approx &  \begin{cases}
\frac{\omega}{d^3}, ~  \omega d/v_F \ll \left(\frac{\omega}{\mu}\right)^{1/3} \ll 1 \\
\frac{\omega^{1/3}}{d^4} , ~ \omega d/v_F \gg 1
\end{cases}
\eeq

At finite temperature $T > \omega$, the spinon self-energy formally diverges because the spinon Green's function is not gauge-invariant.\cite{LeeNagaosa} However, the formal divergence cancels in any gauge-invariant observables, and the $\tz = 3$ scaling can be used to predict the temperature dependence in the $\omega \ll T$ limit as well. For our relaxation time computation, we can write down the spin-spin correlation function by simply using $T^{2/3}$ instead of $\omega^{2/3}$ in the self-energy, which gives
\beq
\mathcal{C}^{~''}_{+-}(\q,\omega) \xrightarrow{T \gg \omega} \frac{\omega}{\sqrt{v_F^2 q^2 + C^2 T^{4/3}}}
\eeq

In the $\omega \ll T$ limit, we have $\omega \coth(\omega/2T) \approx 2T$. So, the relaxation time is given by:
\beq
\frac{1}{T_1}& \approx &  \begin{cases}
\frac{T}{d^3}, ~  T d/v_F \ll \left(\frac{T}{\mu}\right)^{1/3} \ll 1 \\
\frac{T^{1/3}}{d^4} , ~ T d/v_F \gg \left(\frac{T}{\mu}\right)^{1/3}
\end{cases}
\eeq

\subsection{Gapped systems}
\label{app:Integrals}
In this section, we discuss the computations of semi-exact expressions for the relaxation time for different systems. We start off with the case of free non-interacting bosons, when we find that the relaxation time is given by:
\beq
\frac{1}{T_1} &\approx& \int_0^\infty dq \, q^3 e^{- 2 q d} \, \Theta\left( \omega - 2\Delta_s - \frac{q^2}{4m} \right) \nn
&=& \frac{3 - e^{-2Qd}\left[3 + 6 Qd + 6 (Qd)^2 + 4 (Qd)^3 \right]}{d^4}  \Theta\left( \omega - 2\Delta_s \right), \text{ where } Q = \sqrt{4m (\omega - 2 \Delta_s)} \nn
& \approx & \begin{cases}
\frac{Q^4}{4} \Theta\left( \omega - 2\Delta_s \right), ~ Qd \ll 1 \\
\frac{3}{8 d^4} \Theta\left( \omega - 2\Delta_s \right), ~ Qd \gg 1
\end{cases}
\eeq
For non-interacting anyons with statistics parameter $\alpha$, the structure factor considering local two-anyon energy eigenstates, as described in the main text, is given in Ref.~\onlinecite{Sid_PRL2017}.
\beq
\mathcal{C}^{~''}_{+-}(\q,\omega) \propto J_{\alpha}^2(a \sqrt{m(\omega - 2\Delta_s) - q^2/4}) \, \Theta\left( m(\omega - 2\Delta_s) - q^2/4 \right) \nn\approx \left(a \sqrt{m(\omega - 2 \Delta_s) - q^2/4} \right)^{2\alpha}  \Theta\left( m(\omega - 2\Delta_s) - q^2/4 \right) \nn
\eeq
where $a$ is a microscopic lengthscale of the order of lattice spacings. Using the low-energy approximation of the last step, we find that the relaxation time is given by (we set $a = 1$):
\beq
\frac{1}{T_1} &\propto& \int_0^\infty dq \, q^3 e^{- 2 q d} \, \left( \sqrt{4m(\omega - 2 \Delta_s) - q^2} \right)^{2\alpha} \Theta\left( \omega - 2\Delta_s - \frac{q^2}{4m} \right) \nn
&=& \frac{Q^{4 + 2 \alpha}}{2(\alpha + 1)(\alpha + 2)} - \frac{\sqrt{\pi}(Qd)^{5/2+\alpha}}{4z^{4 + 2\alpha}}\Gamma(1+\alpha)\bigg[ 3 I_{5/2 + \alpha}(2Qd) + 2Qd I_{7/2+\alpha}(2Qd) \nn
&& ~~~~~~~~~~~~~~~~~~~~~~~~~~~~~~~~~~~~~~~~~~~~~~~~~~~~~~~~ - 2Qz L_{3/2+\alpha}(2Qz) + 2(1+\alpha)L_{5/2 + \alpha}(2Qz)\bigg] \nn
& \approx & \begin{cases}
\frac{Q^{4+2\alpha}}{2(\alpha + 1)(\alpha + 2)} \Theta\left( \omega - 2\Delta_s \right), ~ Qd \ll 1 \\
\frac{3 Q^{2 \alpha}}{8 d^4} \Theta\left( \omega - 2\Delta_s \right), ~ Qd \gg 1
\end{cases}
\eeq
Above, $I_\nu(x)$ and $L_\nu(x)$ refer to the modified Bessel functions and the modified Struve functions respectively. For fermions with $\alpha = 1$, the expressions are a lot simpler, so we reproduce them below for completeness.
\beq
\frac{1}{T_1} &\propto& \int_0^\infty dq \, q^3 e^{- 2 q d} \, \left( \sqrt{4m(\omega - 2\Delta_s) - q^2} \right)^2 \Theta\left( \omega - 2\Delta_s - \frac{q^2}{4m} \right) \nn
&=& \frac{-15 + 3 Q^2 d^2  + e^{-2Qd}\left[15 + 30 Qd + 27 (Qd)^2 + 14 (Qd)^3 + 4 (Qd)^4 \right]}{8d^6}  \Theta\left( \omega - 2\Delta_s \right), \nn
& = & \begin{cases}
\frac{Q^6}{12} \Theta\left( \omega - 2\Delta_s \right), ~ Qd \ll 1 \\
\frac{3 Q^2}{8 d^4} \Theta\left( \omega - 2\Delta_s \right), ~ Qd \gg 1
\end{cases}
\eeq
Finally, for the case of interacting bosons we have the following form of the dynamic spin structure factor as $T \rightarrow 0$:\cite{Sid_PRL2017,QXS09}
\beq
\mathcal{C}^{~''}_{+-}(\q,\omega) \approx \frac{1}{\left[ \ln\left( 4m(\omega - 2\Delta_s) - q^2/16 b^2\right) + 2 \gamma_E \right]^2 + \pi^2}  \Theta\left( \omega - 2\Delta_s - \frac{q^2}{4m} \right)
\eeq
In principle, the relaxation time can be evaluated numerically using this correlation function. But if we further assume that the range of interaction $Qb \ll 1$ where $b$ is the effective range of interaction,
then we can make analytical progress. In this regime, we can neglect $\gamma_E$ and $\pi$ in comparison to $\ln^2(Qb)$ in the denominator. For $Qd \ll 1$ also ignore the exponential decay factor in the numerator. Then, we have, using $Q = \sqrt{4m(\omega - 2 \Delta_s)}$:
\beq
\frac{1}{T_1} &\propto& \int_0^\infty dq \, \frac{q^3}{\left[ \ln\left[ (Q^2 - q^2)b^2/16 \right] + 2 \gamma_E \right]^2 + \pi^2}  \Theta\left( Q^2 - q^2 \right) \nn 
& \approx & \int_0^\infty dq \, \frac{q^3}{ \ln^2\left[ (Q^2 - q^2)b^2/16 \right] }  \Theta\left( Q^2 - q^2 \right) \nn 
& \approx & \int_0^Q dq \, \frac{q^3}{ \ln^2\left[ (Q^2 - q^2)b^2/16 \right] }   \nn 
& \approx & \frac{1}{b^4} \left[ \frac{1}{2} (Qb/4)^2 \text{Ei}[2 \ln(Qb/4)] - \text{Ei}[4 \ln(Qb/4)] \right]\nn
& \approx & \frac{Q^4}{\ln^2(Qb)}, \text{ when } Qd \ll 1 
\eeq

For $Qd \gg 1$, the exponential factor in the numerator cannot be neglected, but we can still Taylor expand the denominator in powers of $q/Q$ and we find that the dominant contribution to the integral comes from the zeroth order term. Hence, the relaxation time is given by:
\beq
\frac{1}{T_1} &\propto& \int_0^\infty dq \, \frac{q^3 e^{-2qd}}{\ln^2(Qb)} \left[ 1 + \text{O}\left( \frac{q}{Q} \right)^2 \right]  \Theta\left( Q^2 - q^2 \right) \nn
& \approx & \frac{3}{8 d^4 \ln^2(Qb)}
\eeq

\end{widetext}

\bibliographystyle{apsrev4-1}
\bibliography{magNoiseSL}

\end{document}